\documentclass[%
 reprint,
superscriptaddress,
 amsmath,amssymb,
 aps,
floatfix,
]{revtex4-1}
\usepackage{footnote}
\usepackage{algpseudocode}
\usepackage[ruled,lined]{algorithm2e}
\usepackage{graphicx}
\usepackage{epstopdf}
\usepackage{bm}
\usepackage{mathtools}
\usepackage{textcomp}
\usepackage{amsmath}
\usepackage{physics}
\usepackage{float}
\usepackage{siunitx}
\usepackage{amssymb}
\usepackage{float}
\usepackage{steinmetz}
\usepackage{csquotes}
\usepackage{lipsum}
\usepackage{soul,color}
\usepackage{makecell}
\usepackage{bibunits}
\makeatletter
\def\maketitle{
	\@author@finish
	\title@column\titleblock@produce
	\suppressfloats[t]}
\makeatother

\newcommand\tab[1][0.8cm]{\hspace*{#1}}

\begin{document}

\preprint{APS/123-QED}
\title{Optical interaction of the NV$^-$ centre in diamond with a plasmonic metal nanoparticle}

\author{Harini Hapuarachchi}
\email[]{harini.hapuarachchi@rmit.edu.au}

\author{Francesco Campaioli}
\email[]{francesco.campaioli@rmit.edu.au}

\author{Jared H. Cole}
\email[]{jared.cole@rmit.edu.au}

\affiliation{ARC Centre of Excellence in Exciton Science and Chemical and Quantum Physics, School of Science, 
	RMIT University, Melbourne, 3001, Australia}

\date{\today}

\begin{abstract}
We present a rigorous theoretical model for the optical interaction between a nitrogen-vacancy (NV) centre in diamond and a plasmonic metal nanoparticle (MNP), accompanied by a computationally efficient procedure to solve the evolution. The proposed model enables us to successfully explain existing optical emission measurements of NV centres both in the presence and absence of a metal nanoparticle. We show that the NV-plasmon interaction provides a versatile new avenue to enhance and control the optical emission of an NV centre. Changes to the MNP type and size, NV-MNP centre separation, submerging medium permittivity, and NV orientation with respect to the MNP surface can be exploited to improve a plethora of NV centre based nanodevices.

\end{abstract}

\maketitle
\begin{bibunit}[plain]
\subsection*{Introduction} \vspace{-1em} 
Diamond is a unique material with the largest known optical bandgap, high stability, and biocompatibility, providing an attractive platform for many quantum-age technologies \cite{aharonovich2011diamond}. The nitrogen-vacancy (NV) centre in diamond \cite{doherty2013nitrogen} is one of the most photostable solid-state quantum emitters  known to date, operating even at room temperature \cite{doherty2013nitrogen,schietinger2009plasmon}. Diamond quantum technologies based on NV centres are rapidly evolving in areas such as quantum information processing \cite{bermudez2011electron, cai2012long, oberg2019spin, childress2013diamond}, bio-sensing \cite{albrecht2014self, aharonovich2011diamond, hall2012high}, magnetometry \cite{grinolds2013nanoscale, hall2009sensing, hall2010ultrasensitive}, electrometry \cite{dolde2011electric}, thermometry \cite{plakhotnik2014all}, piezometry \cite{lee2017topical}, and lasing \cite{jeske2017stimulated}. NV centres coupled to optical microcavities have recently gained attention as versatile building blocks for applications in quantum information processing and sensing \cite{albrecht2013coupling, aharonovich2011diamond}. 

Due to the presence of strong excitation modes known as localized surface plasmons, plasmonic metal nanoparticles (MNPs) exhibit nanocavity-like optical concentration capabilities, overcoming the half-wavelength size limitation of the conventional microcavities \cite{maier2007plasmonics, premaratne2017theory, hapuarachchi2017cavity}. It is well known that MNPs in the vicinity of a quantum emitter can modify the emission behaviour via changes to the local electric field and the local electromagnetic environment of the emitter \cite{carminati2006radiative, des2008fluorescence, kim1988classical, nisar2021enhanced, hapuarachchi2018exciton, hapuarachchi2019plasmonic, reineck2013distance, schietinger2009plasmon}. Due to their improved light controlling prowess compared to the individual constituents \cite{hapuarachchi2018optoelectronic, gettapola2019control}, nanohybrids comprising MNPs and emitters such as quantum dots have emerged as powerful candidates for a myriad of applications including biosensing \cite{hapuarachchi2019plasmonic}, solar energy harvesting \cite{catchpole2008plasmonic}, quantum information processing \cite{manjavacas2011quantum}, and plasmonic lasing \cite{premaratne2017theory, warnakula2019cavity}.

To the best of our knowledge, the first experimental demonstration of controlled coupling between a diamond nanocrystal containing a single NV centre and a metal nanosphere was done by Schietinger \emph{et al.} through controlled manipulation of the particles with an atomic force microscope \cite {schietinger2009plasmon}. Their experiment demonstrated that hybrid systems comprising metal nanoparticles and NV centres are robust building blocks for novel nanophotonic light sources capable of maintaining the single photon character of the emission, while  remaining stable even at room temperature \cite{schietinger2009plasmon}. As almost all major potential applications of the NV centre rely in some way on its optical interrogation \cite{dolan2014complete}, the work of Schietinger \emph{et al.} provides ample motivation to understand the ability to optically control NV centres using metal nanoparticles further. Despite such promising prospects of improving and controlling NV centre-based nanodevices using metal nanoparticles, the wide array of existing and potential applications based on NV centres, and more than 50 years of NV research \cite{doherty2013nitrogen}, the optical properties of NV-MNP nanohybrids are still not theoretically well understood. 

In this paper, we present a quantum optics-based rigorous theoretical model for the optical interaction of the NV$^-$ centre in diamond (referred to as the \emph{NV centre} hereafter). Our model successfully explains existing optical measurements both in the presence \cite{schietinger2009plasmon} and absence \cite{albrecht2013coupling} of a plasmonic metal nanoparticle. We reveal new insights on the tunability of NV emission spectra using metal nanoparticles, unveiling exciting avenues for future experimental investigations and for the design and control of hybrid NV-plasmonic nanodevices.

\subsection*{Summary of the formalism}
\begin{figure*}[t!]
	\includegraphics[width=\textwidth]{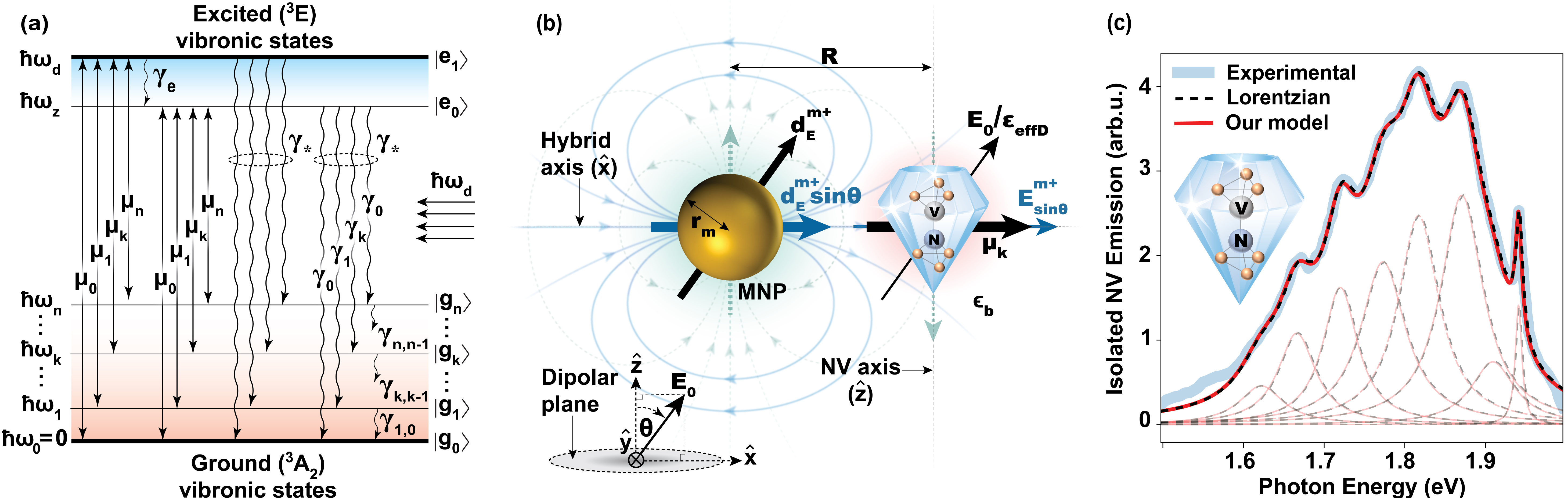}
	\centering
	\caption{(a) Optical model of the NV centre with $n+1$ ground states $\lbrace|g_\text{k}\rangle\rbrace$ with energies $\lbrace\hbar\omega_k\rbrace$ ($k \in \lbrace 0, \hdots, n \rbrace$) and two excited states $|e_0\rangle$ and $|e_1\rangle$. The state $|e_1\rangle$ is a phenomenologically defined upper excited level resonant with the angular frequency of incoming radiation $\omega_\text{d}$. Other parameters are, spontaneous photon emission rate(s) $\gamma_k$, dephasing rate $\gamma_*$, ground state phonon decay rate(s) $\gamma_{k,k-1}$, effective excited state phonon decay rate $\gamma_\text{e}$, and zero phonon line (ZPL) energy $\hbar\omega_\text{z}$. (b) An example dimer setup where the NV-MNP hybrid axis lies along the NV dipolar plane. Optical illumination is polarized along a plane perpendicular to the NV dipolar plane. $\bm{E}_0$ and $\bm{d}_\text{\tiny{E}}^\text{m+}$ denote the positive frequency amplitudes (coefficients of $e^{-i\omega_\text{d}t}$) of the external field and the MNP dipole formed by the external field, respectively. $E_0 \sin\theta/\epsilon_\text{effD}$ and $\bm{E}^\text{m+}_{\sin\theta}$ denote the screened projections of the external field and the MNP dipole response field onto the NV dipolar plane, at the NV location. (c) The experimentally measured emission spectrum of an isolated NV centre in air at room temperature and the fitted Lorentzian sum reported by Albrecht \emph{et al.} in \cite{albrecht2013coupling} are denoted by the solid blue and dashed black lines. The output of our extended NV model in (a) implemented as an open quantum system is plotted in red. Individual emission lines are shown with reduced opacity. Both theoretical and experimental spectra including the fitted Lorentzians and the individual emission bands are normalized by the area of the respective total emission curve. The inset conceptually illustrates the NV centre embedded in a nanodiamond. \label{Fig:Schematic}}
\end{figure*}

In this work, we first combine the insights from Albrecht \emph{et al.} \cite{albrecht2013coupling} and Su \emph{et al.} \cite{su2008towards}, to propose an optical abstraction for the NV centre, as schematically depicted in Fig.\ \ref{Fig:Schematic}(a). We model the NV centre as a multi-level atom with $n+1$ vibronic levels $\lbrace|g_\text{k}\rangle\rbrace$, $k \in \lbrace 0,\hdots,n\rbrace$, in the optical ground state $^3 A_2$. The number of vibronic transitions from $|g_k\rangle$ to the zero phonon level $|g_0\rangle$ is denoted by $k$. The lowest energy level in the optical excited state $^3 E$ is denoted by $|e_0\rangle$. Excited state vibronic levels above $|e_0\rangle$ are represented by an effective upper excited level $|e_1\rangle$ resonant with the energy of incoming radiation. This allows to capture the ability of an NV centre to be optically excited by a photon possessing higher energy than its zero phonon transition $|e_0\rangle\to|g_0\rangle$, for example, by a green photon. Such excitation of the NV centre into $|e_1\rangle$ results in rapid nonradiative decay into the excited band edge $|e_0\rangle$, followed by a radiative transition into one of the ground levels $|g_k\rangle$ emitting a lower energy (red) photon in the $k$th band of the NV emission spectrum. When continuously driven with an optical field, coherent dipolar transitions $|e_i\rangle \leftrightarrow |g_k\rangle$, $i\in\lbrace 0,1\rbrace$, arise in addition to the incoherent decay and dephasing mechanisms within and between the optical states, as depicted in Fig.\ \ref{Fig:Schematic}(a). This abstract optical model can be solved using standard master equation techniques, further details of which are provided in the supplementary material.
	
It has been reported that NV optical transitions are allowed for two degenerate dipoles orthogonal to each other, lying in the plane perpendicular to the NV (symmetry) axis \cite{hui2011polarization, dolan2014complete, albrecht2013coupling}. Several theoretical and experimental studies report the ability of these dipoles to couple equally well to optical fields \cite{fabre2017quantum, alegre2007polarization} resulting in annular patterns of absorption and emission \cite{dolan2014complete, zheng2009orientation} when illuminated with optical fields polarized in the plane perpendicular to the NV axis. This plane will be called the \emph{dipolar plane}, hereafter. Optical electric field polarizations that do not lie along the dipolar plane have been reported to result in polarization angle-dependent anisotropy of the emission intensity \cite{dolan2014complete, alegre2007polarization}, and the observed fluorescence is expected to vanish for field polarizations exactly perpendicular to the dipolar plane, for ideal configurations \cite{dolan2014complete}. In our model, each transition dipole element $\bm{\mu}_k$, $k\in\lbrace0,\hdots,n\rbrace$, in Fig.\ \ref{Fig:Schematic}(a) is assumed to be aligned along the same effective direction defined by the aforementioned degenerate optical dipolar transitions, residing in the NV dipolar plane. 

Optically driven $^3 A_2 \leftrightarrow {^3E}$ transitions are strongly spin preserving \cite{schirhagl2014nitrogen, lee2017topical}, allowing us to ignore the magnetic sublevels and non spin conserving transitions \cite{lee2017topical, schirhagl2014nitrogen, doherty2013nitrogen} as we focus on the NV emission spectra. Generalization of our model to include these effects (for example, in the presence of microwave magnetic fields) is conceptually straightforward, although computationally demanding.

We now consider an NV-MNP dimer illuminated by an external optical electric field with magnitude $E = E_0 e^{-i\omega_\text{d} t} + c.c.$, where $E_0$ is the positive frequency amplitude, $\omega_\text{d}$ is the input frequency, $t$ is time, and $c.c.$ denotes the complex conjugate of the preceding expression. The radius of the MNP is $r_\text{m}$ and it resides at a centre separation $R$ from the NV centre. We obtain the laboratory (static) reference frame NV Hamiltonian below, following the NV optical model in Fig.\ \ref{Fig:Schematic}(a) assuming dipole-dipole type interaction between the MNP and the NV centre:
\begin{multline}\label{Eq:Hamiltonian}
\hat{H}_\text{\tiny{NV}} = \left(\sum_{k=0}^n \hbar\omega_k|g_k\rangle\langle g_k|\right) + \hbar\omega_\text{z}|e_0\rangle\langle e_0| + \hbar\omega_\text{d}|e_1\rangle\langle e_1|\\
-\sum_{k=0}^n\sum_{j=0}^{1}\left(|g_k\rangle\langle e_j| + |e_j\rangle\langle g_k|\right) \mu_k E_\text{tot}.
\end{multline} 
$E_\text{tot}$ is the projection of the total effective electric field experienced by the NV centre on the dipolar plane. In the absence of an MNP, $E_\text{tot}$ is obtainable as the projection of the input field on the NV dipolar plane, screened by the emitter material, diamond. The positive frequency amplitude (coefficient of $e^{-i\omega_\text{d}t}$) of $E_\text{tot}$ in this case would be $\tilde{E}^+_\text{tot} = E_0 \sin{\theta}/\epsilon_\text{\tiny{effD}}$, when $E_0$ forms a polar angle $\theta$ with respect to the NV dipolar plane. The screening factor is given by $\epsilon_\text{\tiny{effD}} = \left(2\epsilon_\text{b} + \epsilon_\text{\tiny{D}}\right)\big/\left(3\epsilon_\text{b}\right)$, where $\epsilon_\text{\tiny{D}}$ and $\epsilon_\text{b}$ are the relative permittivities of diamond and the background medium, respectively. 

For the two special cases where the external field is polarized along the NV dipolar plane either parallel or perpendicular to the hybrid (NV-MNP) axis, we obtain $E_\text{tot}$ as follows,
\begin{align}\label{Eq:E_tot_special}
	&E_\text{tot} = E_\text{tot}^\text{\tiny{(1)+}} + E_\text{tot}^\text{\tiny{(2)+}} + E_\text{tot}^\text{\tiny{(3)+}} + c.c., \;\;\;\text{where,}\\
	&E_\text{tot}^\text{\tiny{(1)+}} = \frac{E_0 e^{-i\omega_\text{d} t} }{\epsilon_\text{\tiny{effD}}}, \;\;\;
	E_\text{tot}^\text{\tiny{(2)+}} = \frac{s_\alpha \alpha(\omega_\text{d}) E_0 e^{-i\omega_\text{d} t} }{\epsilon_\text{\tiny{effD}} R^3} \;\;\;\text{and}\nonumber\\
	&E_\text{tot}^\text{\tiny{(3)+}} = \frac{s_\alpha^2 \alpha(\omega_\text{d})e^{-i\omega_\text{d} t}}{(4\pi\epsilon_0\epsilon_\text{b})\epsilon_\text{\tiny{effD}}^2 R^6}\sum_{j=0}^{1}\sum_{k=0}^{n}\left(\mu_k\tilde{\rho}_{e_j g_k}\right)\nonumber
\end{align}
The orientation parameter $s_\alpha$ takes values 2 or -1 for the cases where the NV dipole orientation is perpendicular (NV$^\perp$MNP) or parallel (NV$^\parallel$MNP) to the MNP surface, respectively. As the plasmonic dipole of a spherical (isotropic) MNP forms along the direction of the effective field incident on it, both MNP and NV dipoles would fall along the hybrid axis in the case where $s_\alpha=2$. Similarly, both dipole types would be perpendicular to the hybrid axis in the case where $s_\alpha=-1$. The slowly varying amplitude (or the rotating frame equivalent) of the off-diagonal NV density matrix element between the $j$th excited state and the $k$th ground state is denoted by $\tilde{\rho}_{e_j g_k}$ and $\epsilon_0$ is the permittivity of free-space.  Polarizability of the MNP (in $\SI{}{\meter\cubed}$ units) at $\omega_\text{d}$ is given by $\alpha(\omega_\text{d})$. For large MNPs, $\alpha(\omega_\text{d})$ is modelled accounting for the finite size effects \cite{colas2012mie, carminati2006radiative}. The recently developed generalized nonlocal optical response (GNOR) theory \cite{raza2015nonlocal} which accounts for the nonlocal effects is used to obtain $\alpha(\omega_\text{d})$ of small MNPs.  Further details of the MNP models used are outlined in the supplementary material. 

The above approach was inspired by a formalism established in the literature in the context of quantum dot-MNP interaction  \cite{zhang2006semiconductor, artuso2008optical, hapuarachchi2018exciton, hapuarachchi2020influence, hapuarachchi2019plasmonic}. Here, we have adapted and extended it to the context of NV-MNP optical interaction, for the first time. From equations (\ref{Eq:Hamiltonian}) and (\ref{Eq:E_tot_special}), it is observable that density matrix elements enter the NV Hamiltonian through the self-feedback field component $E_\text{tot}^\text{\tiny{(3)+}}$ via the MNP, suggesting non-linear evolution of the NV centre (conceptually similar to the case of a quantum dot in the presence of MNPs \cite{zhang2006semiconductor, artuso2008optical, hapuarachchi2018exciton, hapuarachchi2020influence, hapuarachchi2019plasmonic}). 

Using a computationally efficient piecewise superoperator based procedure, we solve the nonlinear evolution of the NV density matrix in a rotating reference frame for the special cases captured by (\ref{Eq:E_tot_special}), and obtain the respective NV emission spectra. In this process, we consider all decoherence mechanisms captured in Fig.\ \ref{Fig:Schematic}(a), as well as both electric field and emission rate modifications caused by the MNP. The derivations, parameters, and the procedures of solution and generation of emission spectra are elaborated in the supplementary material for brevity of the main text. 

Our simulations reveal that, for the entire parameter region explored in this work, the contribution of the positive frequency self-feedback field component of the NV centre ($E_\text{tot}^\text{\tiny{(3)+}}$) is at least five orders of magnitude smaller than the screened sum of the external field and the direct dipole response field of the MNP ($E_\text{tot}^\text{\tiny{(1)+}}+ E_\text{tot}^\text{\tiny{(2)+}}$). In such regions, we can closely approximate $E_\text{tot}$ for any polarization orientation by projecting the screened sum of the external field and the direct dipole response field of the MNP onto the NV dipolar plane, as schematically depicted for a planar example in Fig.\ \ref{Fig:Schematic}(b). The required positive frequency component of the MNP direct dipole response field experienced by the NV centre is obtainable under quasistatic dipole approximation as \cite{griffiths1999introduction}, $\bm{E}^\text{m+}\approx \left[(3 \bm{d}^\text{m+}_\text{\tiny{E}}\cdot\hat{\bm{r}})\hat{\bm{r}} - \bm{d}^\text{m+}_\text{\tiny{E}}\right]/(4\pi\epsilon_0\epsilon_\text{b}\epsilon_\text{\tiny{effD}}R^3)$, where $\hat{\bm{r}}$ is the unit vector of NV centre position relative to the centre of MNP, and $\bm{d}^\text{m+}_\text{\tiny{E}} = (4\pi\epsilon_0\epsilon_\text{b})\alpha(\omega_\text{d}) \bm{E}_0 e^{-i\omega_\text{d} t}$ \cite{maier2007plasmonics}.
\vspace{-0.8em}

\subsection*{Comparison with experimental observations} 
We first validate our extended NV centre model in Fig.\ \ref{Fig:Schematic}(a) by comparing the area normalized emission spectrum generated for an isolated NV centre in air against the experimental measurements and Lorentzian fits by Albrecht \emph{et al.} \cite{albrecht2013coupling}. All three versions of spectra closely overlap as observable in Fig.\ \ref{Fig:Schematic}(c). Throughout this work, we use the set of NV parameters reported by Albrecht \emph{et al.} \cite{albrecht2013coupling} for a single NV centre in a nanodiamond at room temperature. 

We then abstractly replicate the NV-MNP dimer setup of a single NV centre coupled to an MNP by Schietinger \emph{et al.} \cite{schietinger2009plasmon}, and compare their experimental observations to the output of our model. In their dimer-based experiment, a nanodiamond (ND) hosting a single NV centre is kept in close proximity to a $\SI{30}{\nano\meter}$ radius gold nanoparticle (AuNP) on the planar platform of an inverted confocal microscope. The assembled dimer is illuminated with a $\SI{532}{\nano\meter}$ input  laser beam propagating perpendicular to the plane of the confocal microscope platform. Therefore, the electric field oscillations experienced by the dimer occur parallel to the aforementioned plane. The polarization of the beam is controlled (rotated) using a $\lambda/2$ waveplate. The exact position of the NV centre inside the ND is unknown in their experiment and it could vary from 0 to $\approx\SI{40}{\nano\meter}$ from the AuNP surface. They experimentally seek the \emph{optimal configuration} ($\sim$NV$^\perp$MNP arrangement with the shortest achievable NV-AuNP separation) with the aid of atomic force microscope (AFM) based manipulations. This requires the dipolar plane of the NV centre to be perpendicular to the aforementioned confocal microscope platform plane. The top-view of our abstract version of this setup is schematically depicted in Fig.\ \ref{Fig:Schematic}(b), where the externally incident field is polarized along the NV-MNP hybrid axis ($\theta\approx\pi/2$ radians) for the aforementioned optimal configuration in \cite{schietinger2009plasmon}.

\begin{figure}[t!]
	\includegraphics[width=\columnwidth]{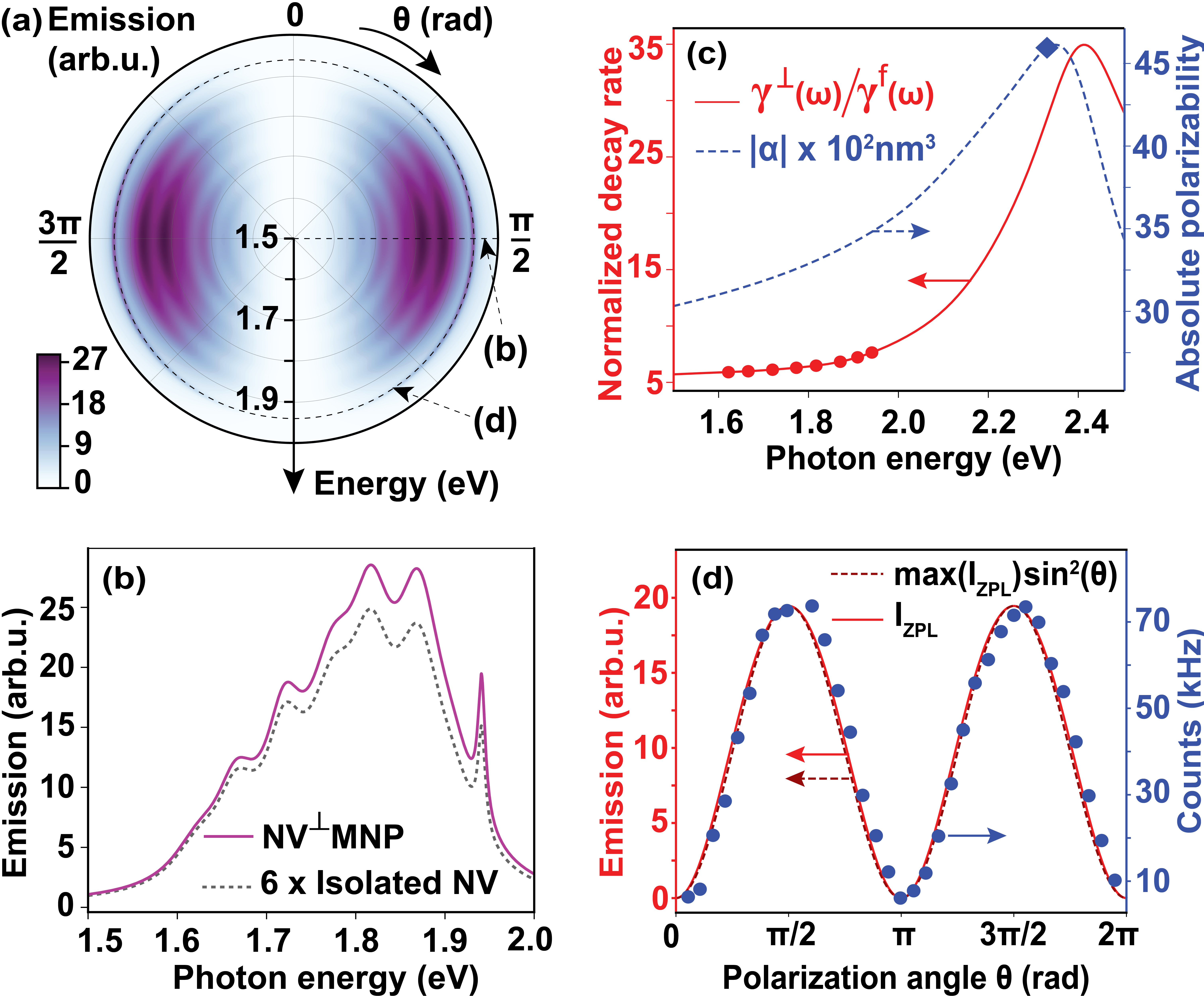} 
	\centering
	\caption{(a) Polar contour plot of NV emission intensities for the setup in Fig.\ \ref{Fig:Schematic}(b), in the presence of a gold nanoparticle of radius $r_\text{m}=\SI{30}{\nano\meter}$ located at a centre separation $R=\SI{38}{\nano\meter}$ from the NV centre in air. (b) NV Emission intensity spectrum of the same NV-MNP setup obtained at $\theta=\pi/2$ radians. Dotted curve depicts the spectrum of the isolated NV centre magnified six times. (c) Modified decay rate spectrum (red curve) for a generic emitter oriented perpendicular to the aforementioned MNP surface, normalized by the respective free-space emission rate $\gamma^f(\omega)$. Red circles capture the normalized decay rates for NV$^\perp$MNP ($\theta=\pi/2$) case for the NV emission band peaks. The dashed blue curve depicts the absolute polarizability of the MNP, and its value at the input laser frequency ($\SI{532}{\nano\meter}$ wavelength) is marked with the blue diamond. (d) Comparison of the calculated ZPL emission intensities against the experimentally detected photon counts in \mbox{\cite{schietinger2009plasmon}}. All emission values are normalized by the area of the isolated NV centre spectrum at $\theta=\pi/2$. \label{Fig:Dimer_Exp}}
\end{figure}

We present emission energy ($\hbar\omega$) and excitation polarization angle ($\theta$) sweeps of the NV emission intensity of the setup in Fig.\ \ref{Fig:Schematic}(b), as a polar surface plot depicted in Fig.\ \ref{Fig:Dimer_Exp}(a). It is observable that all emission sidebands and the zero-phonon line (ZPL) of the NV centre exhibit anisotropy as a function of the polarization angle $\theta$ in the confocal microscope platform plane (perpendicular to the NV dipolar plane). The highest intensity of each emission peak is observed for the $\theta=90^0$ (NV$^\perp$MNP) case. Schietinger \emph{et al.} reported $\sim$6 times enhancement in NV$^-$ emission intensity for the experimentally achieved optimal configuration, compared to the isolated NV$^-$ emission and estimated a quantum efficiency of $\sim$0.78 for the dimer \cite{schietinger2009plasmon}. We obtain a very similar emission enhancement using our model at $R=\SI{38}{\nano\meter}$. In Fig.\ \ref{Fig:Dimer_Exp}(b), we have shown the theoretically obtained total near-field emission intensity of the NV$^-$ centre in the presence of the AuNP, for the NV$^\perp$MNP configuration at this centre separation. The far-field emission spectrum (which can be estimated by scaling the near-field spectrum in  Fig.\ \ref{Fig:Dimer_Exp}(b) by the quantum efficiency of $\sim$0.78 reported in \cite{schietinger2009plasmon}) also exhibits $\sim$6 times ZPL intensity enhancement compared to the isolated NV centre. 

At the same NV-MNP separation $R=\SI{38}{\nano\meter}$, we obtain a total decay rate enhancement (the sum of modified decay rates for all transitions normalized by the sum of all isolated NV transition rates) $\approx 6.5$, and a ZPL decay rate enhancement $\approx 7.6$ as observable in Fig.\ \ref{Fig:Dimer_Exp}(c), using the decay rate modification procedure in \cite{carminati2006radiative} summarized in the supplementary material. These theoretical predictions are in good agreement with the experimentally observed excited state decay rate enhancement $\sim 7.5$ of the aforementioned optimal configuration in \cite{schietinger2009plasmon}. 

We then compare the dependence of NV emission intensity on the polarization angle $\theta$ predicted by our model against the experimentally reported variation in \cite{schietinger2009plasmon}, in Fig.\ \ref{Fig:Dimer_Exp}(d). The theoretically predicted polarization angle dependence of emission intensity is in good agreement with the experimentally observed variation. Both theoretical and experimental emission patterns exhibit an angle dependence closely proportional to $\sin^2\theta$. It was verified that the same pattern as that presented for the ZPL intensity in Fig.\ \ref{Fig:Dimer_Exp}(d) holds for all NV emission band intensities and for their summation, as is also  deducible from Fig.\ \ref{Fig:Dimer_Exp}(a). The imperfect diminishing of the experimentally detected emission intensity for excitation directions along the NV axis ($\theta=0,\pi$ radians) could be attributable to reasons such as imperfect alignment of polarization, sensitivity of linear polarization rotations to background effects, and particle drift \cite{dolan2014complete}.

\begin{figure}[t!]
	\includegraphics[width=\columnwidth]{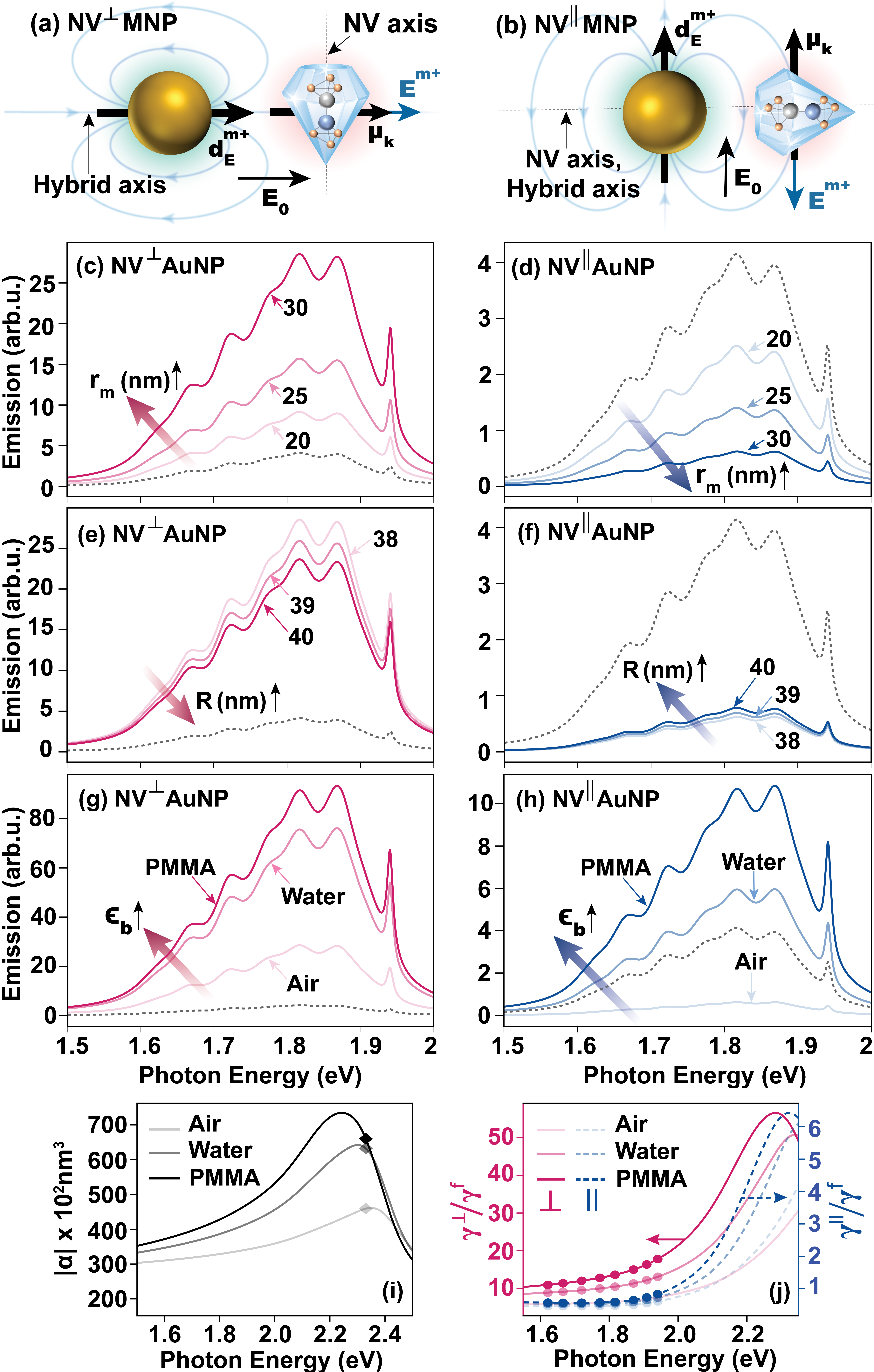}
	\centering
	\caption{(a) Schematic diagram of the NV$^\perp$MNP setup where the external field is polarized along the hybrid axis and NV dipoles are perpendicular to the MNP surface. (b) NV$^\parallel$MNP setup where the external field is polarized perpendicular to the hybrid axis and NV dipoles are parallel to the MNP surface. Subplots (c) and (d) show the variation of total near-field NV emission of the NV$^\perp$AuNP and NV$^\parallel$AuNP dimers in air for different MNP radii at $R=\SI{38}{\nano\meter}$. Subplots (e) and (f) depict a similar analysis for different centre separations ($R$) with MNP radius $r_\text{m}=\SI{30}{\nano\meter}$, in air. Subplots (g) and (h) depict the submerging medium dependence for NV$^\perp$MNP and NV$^\parallel$MNP dimers with $r_\text{m}=\SI{30}{\nano\meter}$ and $R=\SI{38}{\nano\meter}$. Refractive index $n_\text{b} = \sqrt{\epsilon_\text{b}}\approx$ 1, 1.33, and 1.495 for air, water, and PMMA, respectively. Reference (dashed black) curves in subplots (c)-(h) show the isolated NV emission for the respective cases in air. All emission plots are normalized by the area of the respective reference curve. (i) Absolute polarizabilities $|\alpha|$ for $r_\text{m}=\SI{30}{\nano\meter}$ AuNP in different media. Diamond markers depict the values at the illumination frequency. (j) Decay rate modification for generic emitters in different media, oriented $\perp$ and $\parallel$ to the surface of the $r_\text{m}=\SI{30}{\nano\meter}$ AuNP at $R=\SI{38}{\nano\meter}$. Circles depict the values at NV emission peaks. \label{Fig:Large_MNPs}}
\end{figure}

The dipole approximation we use here does not account for effects such as the spatial retardation of incoming radiation, multipolar effects, and any substrate effects. However it is noteworthy that our model effectively captures the essential physics of NV-MNP interaction even with such approximations, in the parameter regime of interest. It is also important to note that we have not fitted the results of Schietinger \emph{et al.} \cite{schietinger2009plasmon}, but rather used the NV parameters reported by Albrecht \emph{et al.} \cite{albrecht2013coupling}, together with the common set of other required parameters outlined in the supplementary material. Yet, our model yields good agreement with the NV-MNP dimer based measurements of Schietinger \emph{et al.} \cite{schietinger2009plasmon}, as evident from the earlier comparisons based on Fig.\ \ref{Fig:Dimer_Exp}. 

\subsection*{Controlling NV emission using MNPs} 
We now investigate the possibility of controlling an NV centre's optical emission using a metal nanoparticle placed at nanoscale proximity. We focus on the NV$^\perp$MNP and NV$^\parallel$MNP setups schematically depicted in Fig.\ \ref{Fig:Large_MNPs}(a) and (b), where the NV dipole orientations are perpendicular and parallel to the MNP surface, respectively. In the NV$^\perp$MNP setup, both MNP and NV dipoles are oriented along the NV-MNP hybrid axis. Therefore, the NV centre experiences an enhanced electric field due to the constructive superposition of the external field and the MNP dipole response field at the NV location, as observable in Fig.\ \ref{Fig:Large_MNPs}(a). Conversely, in the NV$^\parallel$MNP setup, the MNP dipole response field destructively interferes with the external field at the NV location as observable in Fig.\ \ref{Fig:Large_MNPs}(b). This can result in suppression or enhancement of the NV emission intensity (compared to the isolated NV emission) depending on the strength of the MNP dipole response field, as we discuss below.

\begin{figure*}[t!]
	\includegraphics[width=\textwidth]{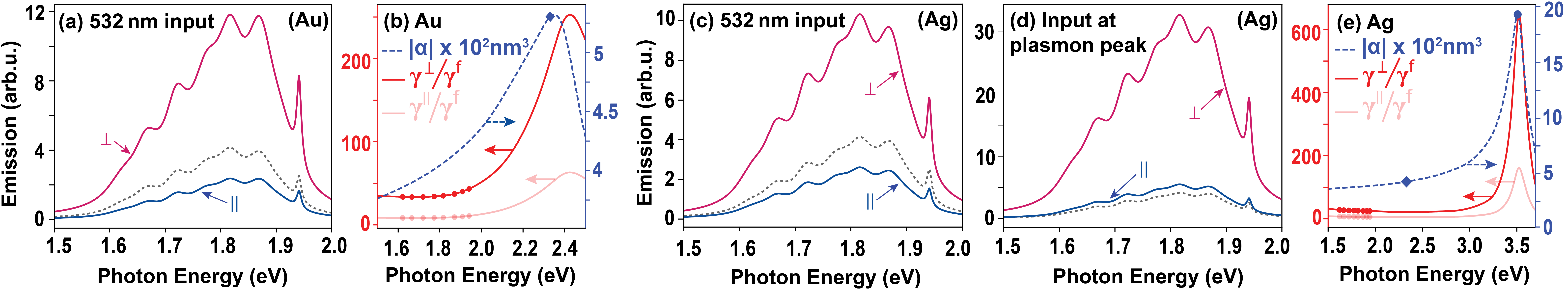}
	\centering
	\caption{(a) Total near-field emission of NV centres in NV$^\perp$AuNP ($\perp$) and NV$^\parallel$AuNP ($\parallel$) dimers in air illuminated at input wavelength $\SI{532}{\nano\meter}$. AuNP radius $r_\text{m}=\SI{7}{\nano\meter}$, NV-AuNP centre separation $R=\SI{12}{\nano\meter}$, submerging medium is air. (b) Absolute polarizability $|\alpha|$ for the $r_\text{m}=\SI{7}{\nano\meter}$ AuNP in air (diamond marker depicts the illumination frequency), and the decay rate modification spectra for generic emitters oriented $\perp$ and $\parallel$ to the AuNP surface, normalized by the respective free-space decay rates $\gamma^\text{f}(\omega)$. The circles depict normalized decay rates at NV emission peaks. Subplots (c) and (d)  depict total near-field emission spectra of NV centres in NV$^\perp$AgNP and NV$^\parallel$AgNP dimers in air, illuminated at $\SI{532}{\nano\meter}$ and at the plasmon resonance frequency of the AgNP, respectively ($r_\text{m}=\SI{7}{\nano\meter}$, $R=\SI{12}{\nano\meter}$). (e) Absolute polarizability of the AgNP (the blue diamond and circle depict the input frequencies of (c) and (d)), and the estimated normalized decay rates for a generic emitter in the presence of the AgNP. Red and pink circles depict the values at NV emission peaks, for  NV$^\perp$AgNP and NV$^\parallel$AgNP orientations, respectively. Reference (dashed) curves in subplots (a), (c), (d)  show the isolated NV emission for the respective cases. All emission plots are normalized by the area of the respective reference curve. \label{Fig:Small_MNPs_Main}}
\end{figure*}

The impact of varying the MNP radius $r_\text{m}$ is captured in Fig.\ \ref{Fig:Large_MNPs}(c) and (d). In Fig.\ \ref{Fig:Large_MNPs}(c), the NV emission intensity increases with $r_\text{m}$ due to the accompanied enhancement of the MNP dipole response field that constructively superposes with the external field, in the NV$^\perp$AuNP setup. Conversely, the NV emission intensity decreases with increasing $r_\text{m}$ in the NV$^\parallel$AuNP setup in Fig.\ \ref{Fig:Large_MNPs}(d), which is indicative of the dominance of the external field over the MNP dipole response field, for all three values of $r_\text{m}$ considered. Here, the reduction in emission intensity occurs due to the gradual increase of MNP dipole response field with increasing $r_\text{m}$, causing the (screened) resultant field experienced by the NV centre to decrease via destructive superposition. 

It is observable from Fig.\ \ref{Fig:Large_MNPs}(e) and (f) that the qualitative impact of reducing the NV-MNP centre separation ($R$) for a fixed MNP radius is similar to that of increasing the MNP radius at a fixed centre separation. This is because both result in increasing the MNP dipole response field at the NV location. Thus, we observe an increase in NV emission intensity with decreasing $R$ for the NV$^\perp$AuNP setup, and vise-versa for the NV$^\parallel$AuNP setup, in the current parameter region.  

We analyse the dependence of steady-state NV emission on the submerging medium permittivity in the presence of a metal nanoparticle in Fig.\ \ref{Fig:Large_MNPs}(g) and (h). It can be observed that NV emission intensity is likely to enhance as the submerging medium permittivity increases, relative to the NV emission intensity observed for the respective dimer in air. The observed enhancement is partially attributable to the larger MNP dipole response field resulting from the increased dipolar polarizability of the MNP depicted in Fig.\ \ref{Fig:Large_MNPs}(i). Increasing the submerging medium permittivity relative to the emitter dielectric permittivity decreases the screening factor $\epsilon_\text{effD}$, resulting in an enhancement of the effective field (hence the effective Rabi frequencies) experienced by the NV centre, contributing to the enhancement of steady state emission intensity. Furthermore, from Fig.\ \ref{Fig:Large_MNPs}(j), we can observe that normalized emission rates for the NV centre's $|e_0\rangle\to|g_k\rangle$ transitions tend to increase with increasing medium permittivity, for both NV$^\perp$AuNP and NV$^\parallel$AuNP configurations. Stronger decay rate enhancements are observed for the NV$^\perp$AuNP case.

We observe similar dependence of NV emission on MNP radius, centre separation, and submerging medium permittivity in the presence of both large and small MNPs. Example results for small MNPs are included in the supplementary material.

We finally explore the impact of illuminating NV-MNP hybrids with input radiation resonant and off-resonant with the plasmonic peak of the MNP in Fig.\ \ref{Fig:Small_MNPs_Main}. To enable comparison between dimers based on relatively high (Au) and low (Ag) dissipation plasmonic materials \cite{stockman2011nanoplasmonics}, we choose $r_\text{m}=\SI{7}{\nano\meter}$ Au and AgNPs in air which can be modelled well within the quasistatic dipole approximation (see supplementary material for details). For $r_\text{m}=\SI{7}{\nano\meter}$ AuNPs in air, the commonly used $\SI{532}{\nano\meter}$ laser resides almost at the plasmon peak frequency as observable in Fig.\ \ref{Fig:Small_MNPs_Main}(b). This results in closely similar emission spectra for illumination at plasmon peak and at $\SI{532}{\nano\meter}$. Therefore, we only present the plots for $\SI{532}{\nano\meter}$ illumination for the AuNP-based dimer, in Fig.\ \ref{Fig:Small_MNPs_Main}(a). However, $\SI{532}{\nano\meter}$ laser frequency and plasmon resonance are far apart for the $\SI{7}{\nano\meter}$ AgNP in air, as observable in Fig.\ \ref{Fig:Small_MNPs_Main}(e). Therefore, we investigate the NV emission spectra for the two illumination possibilities of NV$^\measuredangle$AgNP ($\measuredangle \in \lbrace\perp,\parallel\rbrace$) in Fig.\ \ref{Fig:Small_MNPs_Main} (c) and (d). 
 
Comparing Fig.\ \ref{Fig:Small_MNPs_Main}(a) and (c) reveals that both NV$^\measuredangle$AuNP and NV$^\measuredangle$AgNP ($\measuredangle \in \lbrace\perp,\parallel\rbrace$) display similar levels of NV emission intensity modification, compared to the isolated NV centre emission, at $\SI{532}{\nano\meter}$ illumination. This is because both decay rate modifications at NV emission band peaks and MNP polarizabilities at illumination frequency are comparable for the two cases, as observable in subplots (b) and (e). However, when we illuminate the NV$^\perp$AgNP setup near the plasmon resonance frequency of the AgNP in Fig.\ \ref{Fig:Small_MNPs_Main}(d), a significant emission enhancement compared to the $\SI{532}{\nano\meter}$ case in Fig.\ \ref{Fig:Small_MNPs_Main}(c) is observed due to the larger plasmonic enhancement of the electric field experienced by the NV centre. It is noteworthy that the NV centre experiences an emission enhancement in the NV$^\parallel$AgNP setup, when illuminated at the plasmonic peak of the AgNP in Fig.\ \ref{Fig:Small_MNPs_Main}(d), in contrast to the emission suppression observed for the same setup under $\SI{532}{\nano\meter}$ illumination in Fig.\ \ref{Fig:Small_MNPs_Main}(c). Such enhancement under destructive field superposition (see the NV$^\parallel$MNP schematic in Fig.\ \ref{Fig:Large_MNPs}(b)) indicates strong dominance of the MNP dipole response field over the external field at the NV location, resulting in the formation of a resultant field stronger than the external field, in the opposite direction.

\subsection*{Prospective applications and outlook} 
Our model demonstrates that the photoluminescence of NV centres can be greatly enhanced and controlled using nearby metal nanoparticles. NV-AuNP nanohybrids hold great potential in biomedical applications due to a multitude of reasons: both nanodiamonds and gold nanoparticles are largely inert, biocompatible, and their surfaces can be functionalized with a variety of targeting ligands \cite{albrecht2014self, mieszawska2013multifunctional, aharonovich2011diamond, hui2011polarization,liu2017poly}. The emission of the NV centre readily resides in the near-infrared therapeutic window (650-\SI{900}{\nano\meter}) that exhibits high depths of tissue penetration, and the plasmon resonance \cite{maier2007plasmonics} of AuNPs can be tuned to this region via structural elongation into nanorods \cite{mieszawska2013multifunctional}. Small nanodiamonds suitable for biomarking applications that are about an order of magnitude brighter than traditional red chromophores have been realized \cite{aharonovich2011diamond}. Our model demonstrates that their brightness can be further enhanced, retaining biocompatibility, using AuNPs.

Nanohybrids with low dissipation plasmonic nanoparticles such as silver appear as powerful candidates for nanoscale optoelectronic devices in areas such as quantum information technology \cite{Faraon_2013, childress2013diamond, aharonovich2011diamond} and quantum sensing \cite{wang2014all}. This is due to the large NV emission enhancements achievable by illuminating such hybrids at their plasmon resonance, and the high tunability of the resultant optical field experienced by the NV centre.

Another promising future research avenue is to investigate the conditions under which the nonlinear (density matrix dependent) component of the NV Hamiltonian (\ref{Eq:Hamiltonian}) outweighs the linear component. All of the systems encountered in this work are weakly nonlinear within our choices of metallic plasmonic materials and common submerging media, primarily due to the low coherence between the NV excited and ground states. Investigating and opening pathways to trigger strongly-nonlinear transients in the dynamics of the NV-plasmonic systems can lead to steady-states that depend on the initial state, a condition that can be exploited for enhanced sensing and for discerning NV-plasmonic configurations that cannot be distinguished otherwise.

\subsection*{Conclusion} 
We developed a rigorous theoretical model, validated by experimental results in the literature, for the optical interaction between a coherently illuminated nitrogen-vacancy (NV) centre in diamond and a plasmonic metal nanoparticle (MNP). Using this model, we showed that the impact of varying the MNP size and centre separation on NV emission is highly dependent on the NV dipole orientation with respect to the MNP surface, and that increasing the MNP size is qualitatively analogous to decreasing the NV-MNP centre separation. We further demonstrated the possibilities of significantly enhancing the NV emission intensity by increasing the refractive index of the submerging medium, and by illuminating the NV-MNP hybrid close to the plasmon resonance.

\subsection*{Acknowledgements} 
Authors gratefully acknowledge Andrew D. Greentree and Philipp Reineck for insightful discussions, and the Australian Research Council (grant number CE170100026) for funding. H.H. gratefully acknowledges Dinuka U. Kudavithana for encouragement and support. This work was conducted using the National Computational Infrastructure (NCI), which is supported by the Australian Government.

\providecommand{\noopsort}[1]{}\providecommand{\singleletter}[1]{#1}%

\end{bibunit}
\clearpage
\begin{bibunit}
%
%
\title{Supplementary Material}
\maketitle

\renewcommand{\theequation}{\arabic{equation}}
\setcounter{equation}{0}  

\section*{Theoretical model}
\subsection*{Model overview} \label{Sec:Overview}
In the context of optical interactions, we model the nitrogen-vacancy (NV) centre in diamond as  a multi-level atom with $n+1$ ground states $\lbrace|g_k\rangle\rbrace$, $k\in\lbrace0,\hdots,n\rbrace$, and two excited states $|e_0\rangle$ and $|e_1\rangle$. The number of phononic transitions from $|g_k\rangle$ to the zero phonon level $|g_0\rangle$ is $k$. The lowest energy excited level is denoted by $|e_0\rangle$, while we represent the higher vibronic levels above it by a phenomenologically defined single upper excited level $|e_1\rangle$ resonant with the frequency of incoming radiation. We consider excitation energies well above the zero-phonon transition energy of the NV centre throughout this work. We assume that the NV centre undergoes coherent transitions $|e_i\rangle \leftrightarrow |g_k\rangle$ ($i\in\lbrace 0,1\rbrace$ and $k\in\lbrace0,...,n\rbrace$) upon the incidence of such coherent external radiation. A fast nonradiative (phononic) decay rate $\gamma_\text{e}$ is defined between the two excited states, and dephasing is assumed to occur from both excited states to all ground states at rate $\gamma_*$. The incoherent optical emission transition corresponding to the zero phonon line (ZPL) is $|e_0\rangle\to|g_0\rangle$, whereas the incoherent emission transitions $|e_0\rangle\to|g_k\rangle$ ($k \neq 0$) contribute to the phononic side-bands of the NV optical emission spectrum. The nonradiative transitions between adjacent ground states $|g_k\rangle\to|g_{k-1}\rangle$ are characterized by the phononic decay rates $\gamma_{k,k-1}$ for $k\in\lbrace 1,...,n\rbrace$. The schematic diagram of the NV centre can be found in Fig.\ 1(a) of the main text. 

We assume that the NV centre undergoes optical dipole interactions with a spherical metal nanoparticle (MNP) of radius $r_\text{m}$ placed at a nanoscale centre-separation $R$, upon the incidence of an external coherent electric field, $\bm{E} = E_0 (e^{-i\omega_\text{d} t} + e^{i\omega_\text{d} t}) \hat{\bm{e}}$. The unit vector along the field polarization direction is denoted by $\hat{\bm{e}}$, $\omega_\text{d}$ is the optical frequency, and $t$ is time. Bold fonts denote vector quantities throughout the document. The total effective electric field experienced by the NV centre is altered due to the presence of the MNP. We denote the projection of this total effective field on the \emph{NV dipolar plane} (introduced in the main text) by $\bm{E}_\text{tot}$. Both the diamond nanoparticle hosting the NV centre and the MNP are submerged in a medium of relative permittivity $\epsilon_\text{b}$.

\subsection*{Hamiltonian in the laboratory reference frame} \label{Sec:H_lab}
Setting the energy of the zero-phonon ground state $|g_0\rangle$ to be zero, we can obtain the unperturbed Hamiltonian of the NV centre as follows:
\begin{equation}\label{Eq:H_0}
	\hat{H}_0 = \left(\sum_{k=0}^n \hbar\omega_k|g_k\rangle\langle g_k|\right) + \hbar\omega_\text{z}|e_0\rangle\langle e_0| + \hbar\omega_\text{d}|e_1\rangle\langle e_1|,
\end{equation}
where $\hbar\omega_k$ is the energy of the $k$th ground state phononic level and $\hbar\omega_\text{z}$ is the ZPL energy. The perturbation Hamiltonian component arising due to the coherent dipolar interaction between the NV centre and the total effective field incident on it is obtainable by extending the treatment for two-level emitter-field interaction \cite{fox2006quantum, artuso2012optical, hapuarachchi2018exciton, hapuarachchi2019analysis} as,
\begin{equation}\label{Eq:H_int_vector_form}
	\hat{H}_\text{int} = -\sum_{k=0}^n\sum_{j=0}^{1}\left(|g_k\rangle\langle e_j| + |e_j\rangle\langle g_k|\right) \bm{\mu}_k \cdot \bm{E}_\text{tot}.
\end{equation}
The dipole moment operator element corresponding to the transitions $|e_j\rangle \leftrightarrow |g_k\rangle$ is denoted by $\bm{\mu}_k$. Assuming that $\bm{\mu}_k$ aligns along $\bm{E}_\text{tot}$ we obtain, 
\begin{equation}\label{Eq:H_int}
\hat{H}_\text{int} = -E_\text{tot} \sum_{k=0}^n\sum_{j=0}^{1}\left(|g_k\rangle\langle e_j| + |e_j\rangle\langle g_k|\right) \mu_k.
\end{equation}
The complete form of $E_\text{tot}$ will be elaborated in a later section, for two special cases of interest. The total laboratory frame Hamiltonian of the NV centre under the influence of the externally incident field and the MNP can be obtained as,
\begin{equation}\label{Eq:H_tot}
	\hat{H}_\text{tot} = \hat{H}_0 + \hat{H}_\text{int}.
\end{equation}

\subsection*{Transformation into a rotating reference frame} \label{Sec:H_RF}
To transform the laboratory (static) frame Hamiltonian into a rotating reference frame for the ease of computations, we define a unitary transformation operator,
\begin{equation}\label{Eq:Unitary_operator}
	\hat{U} = e^{i\hat{H}_1 t/\hbar}.
\end{equation}
The Hamiltonian $\hat{H}_1 \approx \hbar\omega_\text{d}\left(|e_0\rangle\langle e_0| + |e_1\rangle\langle e_1|\right)$ is defined in the same eigenbasis as $\hat{H}_0$ with the ground states  $\lbrace|g_0\rangle,\hdots,|g_n\rangle\rbrace$ possessing eigenenergies $\approx 0$ and the excited states $\lbrace|e_0\rangle, |e_1\rangle\rbrace$ possessing eigenenergies $\approx \hbar\omega_\text{d}$. Considering the exponential operator expansion of $e^{i\hat{H}_1 t/\hbar}$ and its adjoint where $|a\rangle$ represents any eigenstate of $\hat{H}_1$, we can obtain the following expressions,
\begin{subequations}
	\begin{align}
		e^{\pm i\hat{H}_1 t/\hbar} |a\rangle &= e^{\pm i\omega_a t} |a\rangle \label{Eq:Exponential expansion_1},\\
		\langle a| e^{\mp i\hat{H}_1 t/\hbar} &= e^{\mp i\omega_a t} \langle a| \label{Eq:Exponential expansion_2},
	\end{align}
\end{subequations}
where $\omega_a$ is the angular frequency of the eigenvalue corresponding to $|a\rangle$.

The laboratory reference frame Hamiltonian (\ref{Eq:H_tot}) can be transformed into the rotating reference frame with the following expression which can be derived using the Schr$\ddot{\text{o}}$dinger equation \cite{jones2012quantum},
\begin{equation}\label{Eq:RF_transformation_eq}
	\hat{H}_\text{\tiny{RF}} = \hat{U}\left(\hat{H}_0+\hat{H}_\text{int}\right)\hat{U}^\dagger + i\hbar\dot{\hat{U}}\hat{U}^\dagger.
\end{equation}
Simplifying (\ref{Eq:RF_transformation_eq}) using (\ref{Eq:Exponential expansion_1}) and (\ref{Eq:Exponential expansion_2}), and applying the rotating wave approximation \cite{blum2012density} (where we discard the fast oscillating terms that average out to zero in the population oscillation timescales of our concern) we obtain,
\begin{multline}\label{Eq:RF_Hamiltonian}
	\hat{H}_\text{\tiny{RF}} \approx \left(\sum_{k=0}^n \hbar\omega_k|g_k\rangle\langle g_k|\right) + \hbar(\omega_\text{z} - \omega_\text{d})|e_0\rangle\langle e_0| \\- \sum_{k=0}^n\sum_{j=0}^{1}\left(\hbar\Omega_k^\text{r}|e_j\rangle\langle g_k| + \hbar\Omega_k^\text{r*}|g_k\rangle\langle e_j| \right),
\end{multline}
where $\Omega_k^\text{r}$ is the effective Rabi frequency at which the NV centre's $|e_j\rangle \leftrightarrow |g_k\rangle$ transition is driven, and $\Omega_k^\text{r*}$ is its complex conjugate. The $k$th Rabi frequency is related to the field incident on the NV dipolar plane such that $\hbar\Omega_k^\text{r} = \mu_k \tilde{E}^+_\text{tot}$, where $\tilde{E}^+_\text{tot}$ denotes the slowly varying positive frequency amplitude of $E_\text{tot} = \tilde{E}^+_\text{tot}e^{-i\omega_\text{d}t}+c.c.$.

\subsection*{Density matrix in the rotating reference frame} \label{Sec:rho_RF}
We first consider the density matrix of the NV centre in the laboratory reference frame, which represents a statistical ensemble of pure states $\lbrace|\psi_l\rangle\rbrace$ (in the same reference frame) occurring with probabilities $\lbrace p_l \rbrace$,
\begin{equation*}\label{Eq:Lab_frame_Density_Matrix}
	\hat{\rho}_\text{\tiny{L}} = \sum_{l} p_l |\psi_l\rangle\langle\psi_l| 
	= \begin{pmatrix}
	\rho_{g_0 g_0} & \hdots & \rho^*_{g_n g_0}  & \rho^*_{e_0 g_0} & \rho^*_{e_1 g_0}\\
	\rho_{g_1 g_0} & \hdots & \rho^*_{g_n g_1} & \rho^*_{e_0 g_1} & \rho^*_{e_1 g_1}\\
	\vdots & \vdots & \vdots & \vdots & \vdots\\
	\rho_{e_0 g_0} & \hdots & \rho_{e_0 g_n}  & \rho_{e_0 e_0} & \rho^*_{e_1 e_0}\\
	\rho_{e_1 g_0} & \hdots & \rho_{e_1 g_n}  & \rho_{e_1 e_0} & \rho_{e_1 e_1}
	\end{pmatrix},
\end{equation*}
where $(^*)$ denotes the complex conjugate. Expanding the above density matrix using the outer products of the NV eigenbasis, we can write,
\begin{equation}\label{Eq:rho_L_expanded}
	\hat{\rho}_\text{\tiny{L}} = \hat{\rho}_\text{\tiny{diag}} + \hat{\rho}_\text{\tiny{off}},
\end{equation}
with
\begin{align*}
	&\hat{\rho}_\text{\tiny{diag}} = \left(\sum_{k=0}^n\rho_{g_k g_k}|g_k\rangle\langle g_k|\right) + \rho_{e_0 e_0}|e_0\rangle\langle e_0|+ \rho_{e_1 e_1}|e_1\rangle\langle e_1|,\\
	&\hat{\rho}_\text{\tiny{off}}\; = \left(\sum_{k=0}^{n} \rho_{e_0 g_k}|e_0\rangle\langle g_k| + \rho_{e_1 g_k}|e_1\rangle\langle g_k|\right) + \rho_{e_1 e_0}|e_1\rangle\langle e_0| \\ & \;\;\;\;\;\; + \left(\sum_{k=1}^{n}\sum_{h=0}^{k-1} \rho_{g_k g_h}|g_k\rangle\langle g_h|\right) + h.c.,
\end{align*}
where $ h.c.$ denotes the Hermitian conjugate of the entire matrix expression that precedes it.

We define a generic state vector transformed into the rotating reference frame $|\tilde{\psi}_l\rangle$ using the unitary operator (\ref{Eq:Unitary_operator}) such that $|\tilde{\psi}_l\rangle = \hat{U}|\psi_l\rangle$, where $|\psi_l\rangle$ is the corresponding state vector in the laboratory (static) reference frame. This enables us to write the density matrix in the rotating reference frame as \cite{steinfeld2013laser, blum2012density},
\begin{equation}\label{eq:Density_matrix_transformation}
\hat{\rho}_\text{\tiny{RF}} = \sum_{l} p_l |\tilde{\psi}_l\rangle\langle\tilde{\psi}| = \sum_{l} p_l \hat{U}|\psi_l\rangle\langle\psi|\hat{U}^\dagger = \hat{U}\hat{\rho}_\text{\tiny{L}}\hat{U}^\dagger.
\end{equation}
The above equation can be simplified using (\ref{Eq:Exponential expansion_1}) and (\ref{Eq:Exponential expansion_1}) to obtain,
\begin{equation}\label{Eq:RF_density_matrix}
	\hat{\rho}_\text{\tiny{RF}} = 
		\begin{pmatrix}
		\rho_{g_0 g_0}         &\hdots& \rho^*_{g_k g_0}       &\hdots& \rho^*_{g_n g_0}          & \tilde{\rho}^*_{e_0 g_0} & \tilde{\rho}^*_{e_1 g_0}  \\
		\vdots                 &\ddots& \vdots                 &\vdots& \vdots                    & \vdots                   & \vdots                    \\
		\rho_{g_k g_0}         &\hdots& \rho_{g_k g_k}         &\hdots& \rho^*_{g_n g_k}          & \tilde{\rho}^*_{e_0 g_k} & \tilde{\rho}^*_{e_1 g_k}  \\
		\vdots                 &\vdots& \vdots                 &\ddots& \vdots                    & \vdots                   & \vdots                    \\
		\rho_{g_n g_0}         &\hdots& \rho_{g_n g_k}         &\hdots& \rho_{g_n g_n}            & \tilde{\rho}^*_{e_0 g_n} & \tilde{\rho}^*_{e_1 g_n}  \\
		\tilde{\rho}_{e_0 g_0} &\hdots& \tilde{\rho}_{e_0 g_k} &\hdots& \tilde{\rho}_{e_0 g_n}    & \rho_{e_0 e_0}           & \rho^*_{e_1 e_0}          \\
		\tilde{\rho}_{e_1 g_0} &\hdots& \tilde{\rho}_{e_1 g_k} &\hdots& \tilde{\rho}_{e_1 g_n}    & \rho_{e_1 e_0}           & \rho_{e_1 e_1}
\end{pmatrix},
\end{equation}
where the elements $\tilde{\rho}_{e_j g_k} = \tilde{\rho}^*_{g_k e_j}$ arise when factoring out the high-frequency time dependence of the coherences between the ground and excited states in the laboratory reference frame density matrix as,
\begin{equation}\label{Eq:High_frequency_factoring}
		\rho_{e_j g_k} = \tilde{\rho}_{e_j g_k} e^{-i\omega_\text{d}t}\;\;\text{for}\;\; k \in \lbrace 0,\hdots,n \rbrace, j\in\lbrace 0,1\rbrace.
\end{equation}
All other elements in (\ref{Eq:RF_density_matrix}) remain the same as those of the laboratory reference frame density matrix $\rho_\text{\tiny{L}}$.

\subsection*{Total effective field experienced by the NV centre} \label{Sec:Effective_electric_field}
Due to each $|e_j\rangle \leftrightarrow |g_k\rangle$ transition in the NV centre, a classically expected oscillating dipole moment of the following form is assumed to be induced (extending the two-level atom based procedure in references \cite{yariv1967quantum, artuso2012optical, hapuarachchi2019analysis, hapuarachchi2020influence}),
\begin{equation}\label{Eq:classically_expected_DM}
	\langle \hat{d}_{jk}\rangle  = d_{jk} = \mu_k\left(\rho_{g_k e_j} + \rho_{e_j g_k}\right),
\end{equation} 
where $\hat{d}_{jk}$ is the respective NV dipole moment operator. The positive frequency components of the dipole moments induced in the MNP in response to each of these oscillating dipole moments in the NV centre can be obtained as \cite{hapuarachchi2020influence},
\begin{equation}\label{Eq:NV_jkth_classical_dipole}
	d_{jk}^\text{m+} = \frac{s_\alpha \alpha(\omega_\text{d}) \mu_k \tilde{\rho}_{e_j g_k} e^{-i\omega_\text{d} t}}{\epsilon_\text{\tiny{effD}} R^3},
\end{equation}
where $\alpha(\omega_\text{d})$ is the polarizability of the MNP  at angular frequency $\omega_\text{d}$, the complete form of which will be presented later. The orientation parameter $s_\alpha=2$ when both NV and MNP dipoles are aligned along the NV-MNP hybrid axis with the NV dipoles ($d_{jk}$) aligned perpendicular to the MNP surface (NV$^\perp$MNP). The parameter $s_\alpha=-1$ when both NV and MNP dipoles are aligned perpendicular to the NV-MNP hybrid axis with the NV dipoles aligned parallel to the MNP surface (NV$^\parallel$MNP). The optical field screening experienced by the NV centre due to the emitter material (diamond) with relative permittivity $\epsilon_\text{\tiny{D}}$ is incorporated using the following screening factor \cite{artuso2012optical};
\begin{equation}
	\epsilon_\text{\tiny{effD}} = (2\epsilon_\text{b} + \epsilon_\text{\tiny{D}})\big/(3\epsilon_\text{b}).
\end{equation}

The positive frequency component of the dipole moment directly induced in the MNP due to the external field, along the same direction, is given by \cite{maier2007plasmonics},
\begin{equation}\label{Eq:MNP_direct_dipole}
	d_\text{\tiny{E}}^\text{m+} = (4\pi\epsilon_0\epsilon_\text{b})\alpha(\omega_\text{d}) E_0 e^{-i\omega_\text{d} t},
\end{equation}
where $\epsilon_0$ is the absolute permittivity of free-space. The positive frequency component of the total dipole moment induced in the MNP due to the external field and the NV transition dipoles can be obtained using (\ref{Eq:NV_jkth_classical_dipole}) and (\ref{Eq:MNP_direct_dipole}) as,
\begin{equation}
	d_\text{tot}^\text{m+} =  d_\text{\tiny{E}}^\text{m+} + \sum_{k=0}^{n}\sum_{j=0}^{1} d_{jk}^\text{m+}.
\end{equation}

We can obtain the positive frequency component of the total effective field incident on the NV centre as the sum of the externally incident field and the total dipole response field of the MNP screened by the diamond lattice (extending the procedure for a two-level emitter implemented in \cite{hapuarachchi2018exciton} and \cite{hapuarachchi2020influence}) as,
\begin{equation}\label{Eq:E_tot_plus}
	E_\text{tot}^+ = \frac{1}{\epsilon_\text{\tiny{effD}}} \left\lbrace E_0 + \frac{s_\alpha d_\text{tot}^\text{m+}}{(4\pi\epsilon_0\epsilon_\text{b}) R^3} \right\rbrace e^{-i\omega_\text{d} t}.
\end{equation}
Expanding the above equation, we can obtain the complete form of the electric field experienced by an NV centre in NV$^\perp$MNP or NV$^\parallel$MNP orientation as,
\begin{widetext}
	\begin{equation}\label{Eq:E_tot}
		E_\text{tot} = \frac{1}{\epsilon_\text{\tiny{effD}}} \left\lbrace E_0 + \frac{s_\alpha \alpha(\omega_\text{d}) E_0}{R^3}  + \frac{s_\alpha^2 \alpha(\omega_\text{d})}{(4\pi\epsilon_0\epsilon_\text{b})\epsilon_\text{\tiny{effD}} R^6}\sum_{j=0}^{1}\sum_{k=0}^{n}\left(\mu_k\tilde{\rho}_{e_j g_k}\right)\right\rbrace e^{-i\omega_\text{d} t} + c.c.,
	\end{equation}
\end{widetext}
where $c.c.$ denotes the complex conjugate of the entire preceding expression.

We can express the positive frequency amplitude of the above field in terms of the Rabi frequency (or its decomposition) for any $|e_j\rangle \leftrightarrow |g_k\rangle$ transition as follows,
\begin{equation}\label{Eq:E_tot_plus_rabi_form}
\tilde{E}_\text{tot}^+ =  \frac{\hbar}{\mu_k}\Omega_k^\text{r} = \frac{\hbar}{\mu_k}\left\lbrace \Omega_k + \eta_k \sum_{j=0}^{1}\sum_{l=0}^{n}\left(\mu_l\tilde{\rho}_{e_j g_l}\right) \right\rbrace,
\end{equation}
where the $\Omega_k$ denotes the Rabi frequency in the absence of coherences (when all $\tilde{\rho}_{e_j g_k} = 0$) and $\eta_k$ is the NV self-interaction coefficient obtainable as, 
\begin{subequations}
	\begin{align}
		\Omega_k &= \frac{\mu_k E_0}{\hbar \epsilon_\text{\tiny{effD}}}\left\lbrace 1 +  \frac{s_\alpha \alpha(\omega_\text{d})}{R^3} \right\rbrace \label{Eq:Omega_k}.\\
		\eta_k &= \frac{\mu_k s_\alpha^2 \alpha(\omega_\text{d})}{(4\pi\epsilon_0\epsilon_\text{b}) \hbar \epsilon_\text{\tiny{effD}}^2 R^6}.
	\end{align}
\end{subequations}
In the absence of the MNP (when $r_\text{m}\to 0$ or $R\to\infty$),
\begin{subequations}
	\begin{align}
	\Omega_k &\to \Omega_k^0 = \frac{\mu_k E_0}{\hbar \epsilon_\text{\tiny{effD}}},\\
	\eta_k &\to \eta_k^0 = 0.
	\end{align}
\end{subequations}

\subsection*{MNP polarization} \label{Sec:MNP_Polarization}
We now focus on the optical response of the metal nanoparticle, incorporated into our model using its dipolar polarizability at angular frequency $\omega_\text{d}$, $\alpha(\omega_\text{d})$. The extensively used format of the MNP polarizability in the literature arises from the solution of Laplace equation for the electric potential $(\nabla^2\phi=0)$ for a homogenous, isotropic sphere embedded in an isotropic, non-absorbing submerging medium upon the incidence of a spatially static (temporally oscillating) electric field \cite{maier2007plasmonics}. The resulting polarizability is given by,
\begin{equation}\label{Eq:LRA_alpha}
\alpha_\text{\tiny{L}}(\omega_\text{d}) = r_\text{m}^3 \frac{\epsilon_\text{m}(\omega_\text{d}) - \epsilon_\text{b}}{\epsilon_\text{m}(\omega_\text{d})+2\epsilon_\text{b}},
\end{equation} 
where $\epsilon_\text{m}(\omega_\text{d})$ is the spatially constant relative permittivity of the metal at frequency $\omega_\text{d}$. It has been shown that this lowest-order (dipolar) full scattering problem under the \emph{quasi-static} and local response approximations (LRA) \cite{raza2015nonlocal} describes the optical properties of nanoparticles of dimensions below $\SI{100}{\nano\meter}$ adequately for many purposes \cite{maier2007plasmonics}. The quasi-static local dipolar polarizability in equation (\ref{Eq:LRA_alpha}) has been further improved in the contexts of large and small MNPs as follows;

\subsubsection*{Large MNPs}
For large MNPs (for example, with diameters $\sim \SI{80}{\nano\meter}$ \cite{des2008fluorescence}), equation (\ref{Eq:LRA_alpha}) which models the MNP as an ideal dipole does not satisfy the optical theorem (energy conservation) \cite{colas2012mie, des2008fluorescence}. This apparent paradox has been overcome by taking the finite-size effects into account, which leads to an effective dipolar polarizability \cite{colas2012mie, des2016plasmonic, carminati2006radiative}, 
\begin{equation}\label{Eq:alpha_eff}
	\alpha_\text{eff}(\omega_\text{d}) = \frac{\alpha_\text{\tiny{L}}(\omega_\text{d})}{\left[1-\frac{2i k_\text{b}^3}{3}\alpha_\text{\tiny{L}}(\omega_\text{d})\right]}.
\end{equation}
Wavenumber of the non-absorbing submerging medium is $k_\text{b} = n_\text{b}k$, where $n_\text{b} = \sqrt{\epsilon_\text{b}}$ is the refractive index of the (non-magnetic) medium and $k=\omega_\text{d}/c$ is the free-space wavenumber ($c$ is the speed of light). The effective polarizability $\alpha_\text{eff}$ accounts for the radiative reaction (impact of finite size) of the MNP which microscopically originates from radiation emitted by the charge oscillations induced inside the nanoparticle by the external field \cite{colas2012mie}.

\subsubsection*{Small MNPs}
Size dependent plasmon resonance shifts and linewidth broadening phenomena that cannot be captured using the local optical polarizability (\ref{Eq:LRA_alpha}) has been observed in recent plasmonic experiments involving MNPs of diameter $\lesssim \SI{20}{\nano\meter}$ \cite{raza2015nonlocal}. These effects arise due to a nanoscale physical mechanism beyond classical electrodynamics known as the nonlocal response \cite{christensen2014nonlocal}. The \emph{generalized nonlocal optical response} (GNOR) theory that accounts for both electron pressure and electron diffusion effects in such small MNPs was recently introduced by Mortensen and Raza \emph{et al} \cite{mortensen2014generalized, raza2015nonlocal}. The GNOR theory introduces a nonlocal correction ($\delta_\text{\tiny{NL}}$) to the LRA based dipolar polarizability in (\ref{Eq:LRA_alpha}) as follows \cite{raza2015nonlocal}, 
\begin{equation}\label{Eq:alpha_NL}
	\alpha_\text{\tiny{NL}}(\omega_\text{d}) = r_\text{m}^3 \frac{\epsilon_\text{m}(\omega_\text{d}) - \epsilon_\text{b}\left[1+\delta_\text{\tiny{NL}}(\omega_\text{d})\right]}{\epsilon_\text{m}(\omega_\text{d})+2\epsilon_{b}\left[1+\delta_\text{\tiny{NL}}(\omega_\text{d})\right]}.
\end{equation}
The nonlocal correction is given by,
\begin{equation}\label{Eq:delta_NL}
	\delta_\text{\tiny{NL}}(\omega_\text{d}) = \frac{\left[\epsilon_\text{m}(\omega_\text{d}) - \epsilon_\text{core}(\omega_\text{d})\right]j_1(k_\text{\tiny{NL}}(\omega_\text{d})r_m)}{\epsilon_\text{core}(\omega_\text{d}) k_\text{\tiny{NL}}(\omega_\text{d})r_m j_1' (k_\text{\tiny{NL}}(\omega_\text{d})r_m)},
\end{equation}
where $k_\text{\tiny{NL}}$ is the longitudinal wavenumber in the GNOR model obtainable as \cite{raza2015nonlocal},
\begin{equation}
	k_\text{\tiny{NL}}^2(\omega_\text{d}) = \frac{\omega_\text{d}(\omega_\text{d}+i\Gamma_\text{m})\epsilon_\text{m}(\omega_\text{d})}{\epsilon_\text{core}(\omega_\text{d})\left[\beta^2+D_\text{m}(\Gamma_\text{m} - i\omega_\text{d})\right]}. 
\end{equation}
In the above equations, the response of bound electrons, bulk plasmon damping rate, and electron diffusion constant of the MNP are denoted by $\epsilon_\text{core}(\omega_\text{d})$, $\Gamma_\text{m}$, and $D_\text{m}$, respectively. In the high-frequency limit where $\omega_\text{d}\gg \Gamma_\text{m}$, $\beta^2 = (3/5)v_\text{\tiny{F}}^2$ where $v_\text{\tiny{F}}$ denotes the Fermi velocity of the MNP. 

We use $\alpha(\omega_\text{d}) = \alpha_\text{eff}(\omega_\text{d})$ for large nanoparticles, and $\alpha(\omega_\text{d})=\alpha_\text{\tiny{NL}}(\omega_\text{d})$ for small nanoparticles when generating our results using the equations presented in earlier sections.

\subsection*{MNP-induced NV decay rate modifications} \label{Sec:Decay_mod}

It has been shown that the rate of radiative emission is not an inherent property of emitters such as quantum dots and NV centres \cite{schietinger2009plasmon, pelton2015modified}. This rate is rather determined by the interaction between the emitter and its local electromagnetic environment \cite{pelton2015modified}, which in this case comprises the metal nanoparticle. To capture such modifications we incorporate the equations outlined in the following sections into our model.

\subsubsection*{Large MNPs}
To estimate the decay rate modification experienced by the $k$th NV emission transition for the cases of normal ($\perp$) and tangential ($\parallel$) emitter dipole orientations with respect to the surface of an adjacent large MNP, we use the following equations derived by Carminati \emph{et al.} where the MNP is treated within the dipole approximation, while accounting for the finite size effects \cite{carminati2006radiative},
\begin{widetext}
\begin{subequations}\label{Eq:Decay_near_large_MNPs}
	\begin{align}
		\frac{\gamma_k^\perp}{\gamma_k^\text{f}} &\approx n_\text{b}\left\lbrace 1 + 6 k_\text{b}^3 \mathrm{Im}\left[\alpha_\text{eff}(\omega_k) e^{2ik_\text{b}R}\left(\frac{-1}{(k_\text{b}R)^4} + \frac{2}{i(k_\text{b}R)^5} + \frac{1}{(k_\text{b}R)^6}\right)\right]\right\rbrace, \\
		\frac{\gamma_k^\parallel}{\gamma_k^\text{f}} &\approx n_\text{b}\left\lbrace 1 + \frac{3}{2}k_\text{b}^3\mathrm{Im}\left[\alpha_\text{eff}(\omega_k)e^{2ik_\text{b}R}\left(\frac{1}{(k_\text{b}R)^2} - \frac{2}{i(k_\text{b}R)^3} - \frac{3}{(k_\text{b}R)^4} + \frac{2}{i(k_\text{b}R)^5} +\frac{1}{(k_\text{b}R)^6}\right)\right] \right\rbrace.
	\end{align}
\end{subequations}
\end{widetext}
The above equations consider an emitter in a medium with refractive index $n_\text{b}$. The free-space decay rate of the emission transition considered is denoted by $\gamma_k^\text{f}$, and $\omega_k$ is the angular frequency of the same transition.  

\subsubsection*{Small MNPs}
For small MNPs submerged in low refractive index media $\alpha_\text{eff}\approx\alpha_\text{\tiny{L}}$. Using equations (\ref{Eq:LRA_alpha}) and (\ref{Eq:alpha_eff}) together with (interpolated) MNP permittivity data from Johnson and Christy's tabulations \cite{johnson1972optical}, we verified this claim for small gold nanoparticles with radii $r_\text{m}\lesssim\SI{7}{\nano\meter}$ in media with refractive index $n_\text{b} \lesssim 1.5$, and for small silver nanoparticles with $r_\text{m}\lesssim\SI{7}{\nano\meter}$ in air. By incorporating the GNOR based nonlocal correction to the decay rate modification equations derived by des Francs \emph{et al.} \cite{des2016plasmonic, colas2012mie} for $k_\text{b}R \ll 1$ we obtain,
\begin{subequations}\label{Eq:Decay_near_small_MNPs}
	\begin{align}
		\frac{\gamma_k^\perp}{\gamma_k^\text{f}} &\approx \frac{6 n_\text{b} \mathrm{Im}\left[\alpha_\text{\tiny{NL}}(\omega_k)\right]}{k_\text{b}^3 R^6},\\
		\frac{\gamma_k^\parallel}{\gamma_k^\text{f}} &\approx \frac{3 n_\text{b} \mathrm{Im}\left[\alpha_\text{\tiny{NL}}(\omega_k)\right]}{2k_\text{b}^3 R^6}.
	\end{align} 
\end{subequations}
We recently used the above equations to successfully describe the decay rate modification of quantum dots near small MNPs, at $\sim$$\SI{10}{\nano\meter}$ surface separations \cite{nisar2021enhanced}. 

\subsection*{NV centre as an open quantum system} \label{Sec:open_quantum_system}
The Hamiltonian of the NV centre optically coupled to the externally incident field and the MNP dipole response field represents a closed quantum system where the impact of the environment (bath) is yet to be taken into account. It couples with the environment resulting in an open quantum system with irreversible dynamics. We estimate the evolution of the density matrix of an open quantum system weakly coupled to a Markovian (memoryless) bath using the following master equation \cite{breuer2002theory},
\begin{multline}\label{Eq:Density_matrix_master_eq}
\dot{\hat{\rho}}_\text{\tiny{RF}} = -\frac{i}{\hbar}[\hat{H}_\text{\tiny{RF}}, \hat{\rho}_\text{\tiny{RF}}] \\+ \sum_{x}\Gamma_x[\hat{L}_x^{\phantom{\dagger}}\hat{\rho}_\text{\tiny{RF}}\hat{L}_x^\dagger -\frac{1}{2}\lbrace \hat{L}_x^\dagger \hat{L}_x^{\phantom{\dagger}}, \hat{\rho}_\text{\tiny{RF}}\rbrace],
\end{multline}
where $\hat{L}_x$ is the Lindblad or collapse operator corresponding to the $x$th decoherence channel with characteristic decoherence rate $\Gamma_x$. The mathematical operators $[\cdot,\cdot]$ and $\lbrace\cdot,\cdot\rbrace$ denote the commutator and anti-commutator of the operands. The decoherence channels and the corresponding rates considered in our extended NV centre model schematically depicted in Fig.\ 1(a) of the main text are as follows:

\begin{flushleft}
	For each optical decay transition $|e_0\rangle \to |g_k\rangle$:\\
	$\Gamma_x = \gamma_k$, for $k\in\lbrace 0,\hdots,n\rbrace$\\
	$\hat{L}_x = \hat{\sigma}_k = |g_k\rangle\langle e_0|$
\end{flushleft}

\begin{flushleft}
	For each phononic decay transition $|g_k\rangle \to |g_{k-1}\rangle$:\\
	$\Gamma_x = \gamma_{k,k-1}$, for $k\in\lbrace 1,\hdots,n\rbrace$\\
	$\hat{L}_x = |g_{k-1}\rangle\langle g_k|$
\end{flushleft}

\begin{flushleft}
	Nonradiative decay in the excited state $|e_1\rangle \to |e_0\rangle$:\\
	$\Gamma_x = \gamma_\text{e}$\\
	$\hat{L}_x = |e_0\rangle\langle e_1|$
\end{flushleft}

\begin{flushleft}
	Dephasing from excited to ground states:\\
	$\Gamma_x = \gamma_*$\\
	$\hat{L}_x = |e_0\rangle\langle e_0|+|e_1\rangle\langle e_1|$
\end{flushleft}

\subsection*{Emission intensity spectrum}
This section outlines how we can utilize the steady state density matrix obtainable by solving (\ref{Eq:Density_matrix_master_eq}) to estimate the emission intensity spectrum of the NV centre. The free-space fluorescence or emitted power spectrum $S_f(\omega)$ of a generic two-level emitter in a stationary state can be calculated using its emission correlation function in the following form \cite{carmichael1999statistical, breuer2002theory, meystre2007elements, nation2011qutip},
\begin{equation}\label{Eq:TLA_emission_spectrum}
	S_f(\omega) = f(\bm{r})\int_{-\infty}^{\infty}d\tau e^{-i\omega\tau}\langle\hat{\sigma}^\dagger(\tau)\hat{\sigma}(0)\rangle_\text{ss},
\end{equation}
employing the homogeneity in time of the stationary correlation function. In the above equation, $\hat{\sigma} = |g\rangle\langle e|$ denotes the emission operator from an excited state $|e\rangle$ to a ground state $|g\rangle$, $\omega$ denotes angular frequency, and $\langle\cdot\rangle_\text{ss}$ is the expectation calculated using the steady state density matrix. The coefficient $f(\bm{r})$ is a geometrical factor defined such that $f(\bm{r})\propto \omega_\text{eg}\gamma$, where $\omega_\text{eg}$ and $\gamma$ denote the emitter resonance frequency and the free-space decay rate, respectively. The vector $\bm{r}$ measures positions with respect to an origin at the location of the emitter \cite{carmichael1999statistical}. We can estimate the photon emission intensity by normalizing the power spectrum in (\ref{Eq:TLA_emission_spectrum}) by the emitted photon energy $\approx\hbar\omega_\text{eg}$. Assuming emission behaviour analogous to the above generic two-level case for each $|e_0\rangle \to |g_k\rangle$ transition, and summing up the intensity spectra resulting from all such transitions, we estimate the total photon emission intensity spectrum of the NV centre as,
\begin{equation}\label{Eq:Total_emission_intensity}
	I_\text{tot}(\omega) \propto \sum_{k=0}^{n} \gamma_k \int_{-\infty}^{\infty}d\tau e^{-i\omega\tau}\langle\hat{\sigma}_k^\dagger(\tau)\hat{\sigma}_k(0)\rangle_\text{ss},
\end{equation}
where $\hat{\sigma}_k = |e_0\rangle\langle g_k|$. This expression is validated by comparison against the experimentally observed NV emission intensity spectra in Fig. 1(c) in the main text.

\section*{Numerical Implementation}

\subsection*{The piecewise superoperator method}
To numerically simulate the NV centre's emission behaviour, we first need to solve the master equation (\ref{Eq:Density_matrix_master_eq}) for its steady state density matrix. The common procedure used in the literature \cite{hatef2012plasmonic, artuso2012optical} when numerically solving similar emitter-MNP systems is the element-wise decomposition of the master equation into a set of coupled differential equations followed by the use of differential equation solvers such as \emph{Runge-Kutta} \cite{noye2000computational} implementations readily available in Matlab and Python. Due to the evolution timescales involved in our problem, such procedures take extremely long computational times to reach the steady state. To solve the problem in hand within much shorter computational times, we propose the following piecewise superoperator method.

We first decompose the Hamiltonian $\hat{H}_\text{\tiny{RF}}$ in (\ref{Eq:RF_Hamiltonian}) experiencing the Rabi frequencies in (\ref{Eq:E_tot_plus_rabi_form}) into linear and nonlinear components as $\hat{H}_\text{\tiny{RF}} \approx \hat{H}_\text{lin} + \hat{H}_\text{nl}$, where the linear part is,
\begin{multline}\label{Eq:H_lin}
		H_\text{lin} = \left(\sum_{k=0}^{n} \hbar\omega_k|g_k\rangle\langle g_k|\right) + \hbar(\omega_\text{z}-\omega_\text{d})|e_0\rangle\langle e_0| \\
	-\sum_{k=0}^{n}\sum_{j=0}^{1}\hbar\Omega_k |e_j\rangle\langle g_k| + \hbar\Omega_k^*|g_k\rangle\langle e_j|.
\end{multline}
The nonlinear part is,
\begin{equation}\label{Eq:H_nl}
	H_\text{nl} = -\sum_{k=0}^{n}\sum_{j=0}^{1}\hbar\text{nl}^\text{\tiny{coeff}}_k |e_j\rangle\langle g_k| + \hbar\text{nl}^\text{\tiny{coeff}*}_k|g_k\rangle\langle e_j|,
\end{equation}
where ($^*$) denotes the complex conjugate and,
\begin{equation}\label{Eq:nl_coeff}
	\text{nl}^\text{\tiny{coeff}}_k = \eta_k \sum_{j=0}^{1}\sum_{l=0}^{n}\left(\mu_l\tilde{\rho}_{e_j g_l}\right).
\end{equation}
Then we judiciously insert the identity operator $\hat{I}$ into the master equation (\ref{Eq:Density_matrix_master_eq}) as,
\begin{multline}\label{Eq:Master_eq_with_identity}
	\hat{I}\dot{\hat{\rho}}\hat{I} = -\frac{i}{\hbar}(\hat{H}_\text{\tiny{RF}} \hat{\rho}_\text{\tiny{RF}}\hat{I} - \hat{I}\hat{\rho}_\text{\tiny{RF}}\hat{H}_\text{\tiny{RF}} ) \\+ \sum_{x}\Gamma_x^{\phantom{\dagger}}[L_x^{\phantom{\dagger}}\hat{\rho}_\text{\tiny{RF}}L_x^\dagger -\frac{1}{2}(L_x^\dagger L_x^{\phantom{\dagger}}\hat{\rho}_\text{\tiny{RF}}\hat{I} + \hat{I}\hat{\rho}_\text{\tiny{RF}}L_x^\dagger L_x^{\phantom{\dagger}})],
\end{multline}
Using the following vector identity in the column-ordered form \cite{barnett1990matrices,byron2012mathematics},
\begin{equation}
	\text{vec}({\hat{A}\hat{X}\hat{B}}) = (\hat{B}^T\otimes\hat{A})\text{vec}(\hat{X}),
\end{equation}
we can obtain the following superoperator form of the master equation,
\begin{equation}\label{Eq:Superoperator form}
	\dot{\vec{\rho}}_\text{\tiny{RF}} = \hat{\mathcal{L}}\vec{\rho}_\text{\tiny{RF}}.
\end{equation}
The vectorised density matrix in the column-ordered form is given by $\vec{\rho}_\text{\tiny{RF}} = \text{vec}(\rho_\text{\tiny{RF}})$, and $\hat{\mathcal{L}} = \hat{\mathcal{L}}_\text{lin} + \hat{\mathcal{L}}_\text{nl}$ is the superoperator (Liouvillian) decomposed into its linear and nonlinear components, where the linear part is,
\begin{align}\label{Eq:Lin_superop}
	&\hat{\mathcal{L}}_\text{lin} = -\frac{i}{\hbar}\left[\hat{I}\otimes\hat{H}_\text{lin} - \hat{H}_\text{lin}^T\otimes\hat{I}\right]+\\
	&\sum_{x}\Gamma_x \left\lbrace (L_x^*\otimes L_x^{\phantom{*}}) - \frac{1}{2} \left[(L_x^\dagger\nonumber L_x^{\phantom{\dagger}})^\text{T}\otimes\hat{I} + (\hat{I}\otimes L_x^\dagger L_x^{\phantom{\dagger}}) \right]\right\rbrace,
\end{align}
and the nonlinear part (that depends on elements of $\hat{\rho}_\text{\tiny{RF}}$) is given by,
\begin{equation}\label{Eq:nl_superop}
	\hat{\mathcal{L}}_\text{nl}(\hat{\rho}_\text{\tiny{RF}}) = -\frac{i}{\hbar}\left[\hat{I}\otimes\hat{H}_\text{nl} - \hat{H}_\text{nl}^T\otimes\hat{I}\right].
\end{equation}
In the absence of non-linearities (when $\hat{\mathcal{L}}=\hat{\mathcal{L}}_\text{lin}$ is independent of both density matrix elements and time), the solution to (\ref{Eq:Superoperator form}) takes the form,
\begin{equation}
	\vec{\rho}_\text{\tiny{RF}}(t) = e^{\hat{\mathcal{L}}_\text{lin}t}\vec{\rho}_\text{\tiny{RF}}(0).
\end{equation}
That is, $e^{\hat{\mathcal{L}}_\text{lin}t}$ propagates a linear system from the initial state to the state at time $t$. To solve the nonlinear problem in hand, we subdivide the total propagation timescale into small (adaptive) time-steps $\delta t$ within each of which the system is assumed to exhibit piecewise linear behaviour of the form,

\begin{align}\label{Eq:piecewise_propagation}
	&\vec{\rho}_\text{\tiny{RF}}(t+\delta t) \approx e^{\hat{\mathcal{L}}(t)\delta t}\vec{\rho}_\text{\tiny{RF}}(t),\\
	&\text{where,}\;\;\;\hat{\mathcal{L}}(t) = \hat{\mathcal{L}}_\text{lin} + \hat{\mathcal{L}}_\text{nl}(\hat{\rho}_\text{\tiny{RF}}(t)). \nonumber
\end{align}
The nonlinear piecewise evolution can be implemented using the \emph{liouvillian}() function of the Quantum Toolbox in Python (QuTiP) \cite{johansson2012qutip} as outlined in Algorithm \ref{Alg:Piecewise_superop}.

\begin{algorithm}\label{Alg:Piecewise_superop}
	\caption{Piecewise superoperator evolution}
	\vspace{0.4em}
	\textbf{input:}\\
	Final times for evolution regions, $\text{tf}_\text{list} = [T_1, T_2, T_3]$\\
	Adaptive time-steps, $\delta t_\text{list} = [\delta t_1, \delta t_2, \delta t_3]$\\
	Linear Hamiltonian, $\hat{H}_\text{lin}$\\
	List of collapse operators with rates, $c_\text{ops} = \lbrace \sqrt{\Gamma_x}\hat{L}_x \rbrace$\\
	List of $\eta_k$ values, $\eta_\text{list} = [\eta_0,\hdots,\eta_n]$\\
	List of $\mu_l$ values, $\mu_\text{list} = [\mu_0,\hdots,\mu_n]$\\
	Initial state, $\hat{\rho}_0=\hat{\rho}_\text{\tiny{RF}}(0)$\\ 
	Initial time, $t_0 = 0$\\
	\vspace{0.4em}
	\textbf{output:}\\
	List of evolution times, $t_\text{list}$\\
	List of evolved states, $\hat{\rho}_\text{list}$\\
	\vspace{0.4em}
	\textbf{begin:}\\
	Initialize empty lists $t_\text{list}, \hat{\rho}_\text{list}$\\
	Initialize current state and time $\hat{\rho}_\text{\tiny{RF}}=\hat{\rho}_0$, $t=t_0$\\ 
	$t_\text{list}.\text{append}(t)$\\
	$\hat{\rho}_\text{list}.\text{append}(\hat{\rho}_\text{\tiny{RF}})$\\
	$\hat{\mathcal{L}}_\text{lin} = liouvillian(\hat{H}_\text{lin}, c_\text{ops})$\\
	\vspace{0.4em}
	\textbf{for} each $T_i$ in $\text{tf}_\text{list}$:\\
	\tab\textbf{while}{ $t<\text{tf}_\text{list}[i]$}:\\
	\tab\tab $t = t + \delta t_i$\\
	\tab\tab Build $\hat{H}_\text{nl}$ using $\eta_\text{list}$, $\mu_\text{list}$ and current $\hat{\rho}_\text{\tiny{RF}}$\\
	\tab\tab $\hat{\mathcal{L}} = \hat{\mathcal{L}}_\text{lin} + liouvillian(\hat{H}_\text{nl}, [\;])$\\
	\tab\tab $\hat{\mathcal{P}}_i = e^{\hat{\mathcal{L}}\delta t_i}$\\
	\tab\tab $\text{vectorized}(\hat{\rho}_\text{\tiny{RF}}) = \hat{\mathcal{P}}_i \times\text{vectorized}(\hat{\rho}_\text{\tiny{RF}})$
	\tab\tab $\hat{\rho}_\text{\tiny{RF}} = vector\_to\_operator\left(\text{vectorized}(\hat{\rho}_\text{\tiny{RF}})\right)$
	\tab\tab $t_\text{list}.$append($t$) \\ 
	\tab\tab $\hat{\rho}_\text{list}.$append($\hat{\rho}_\text{\tiny{RF}}$) \\
	\tab\textbf{end}\\
\textbf{end} 
\end{algorithm}

When implementing the algorithm presented above, we decomposed the total evolution time ($\approx$\SI{300}{\nano\second}) into three regions with final times $[T_1=10^{-4}T, T_2=10^{-3}T, T_3=10^3T]$, and time	 steps $[\delta t_1 = 10^{-6}T, \delta t_2 =10^{-5}T, \delta t_3 =5T]$ for the three regions, respectively. A characteristic time $T=2\pi/|\Omega_0^\text{f}|$ was defined, where $\Omega_0^\text{f} = \mu_0 E_0/(\hbar\epsilon_\text{effD})$ in air, obtained using parameters outlined in the next section. 

The density matrix evolution resulting from the proposed piecewise superoperator method was verified against the results of (4-5th order) \emph{Runge Kutta} implementation in Python for the parameter region of our concern. The newly proposed method together with the carefully chosen adaptive time-steps reduces the computational time taken by a single evolution of (\ref{Eq:Density_matrix_master_eq}) to reach the steady state from several days on a supercomputer (for the conventional \emph{Runge-Kutta} solving) to a few seconds on a generic computer.

\begin{table}[t!]
	\begin{center}
		\begin{tabular}{ |c|c|c|c|c| } 
			\hline
			k & \makecell{$A_k$ \\(arb.u)} & \makecell{$\hbar\omega_k$ \\ $(\SI{}{\milli\electronvolt})$} & \makecell{$\gamma_k^\text{\tiny{f}}$\\$(\SI{}{\mega\hertz})$} & \makecell{$\gamma_{k,k-1}$\\$(\SI{}{\tera\hertz}$)} \\ 
			\hline
			0 & 1520  & 0     & 0.69  & -  \\ 
			1 & 5260  & 31.8  & 2.42  & 85 \\ 
			2 & 18600 & 70.3  & 8.57  & 82 \\
			3 & 16400 & 124   & 7.57  & 79 \\ 
			4 & 14000 & 168   & 6.46  & 88 \\ 
			5 & 9180  & 221   & 4.23  & 65 \\
			6 & 6570  & 275   & 3.03  & 71 \\ 
			7 & 3270  & 319   & 1.51  & 86 \\
			\hline
		\end{tabular}
		\caption{Room temperature NV parameters from \cite{albrecht2013coupling}.}
		\label{Table:Params_from_Roland}
	\end{center}
\end{table}

\begin{table}[t!]
	\begin{center}
		\begin{tabular}{ |c|c|c|c|c| } 
			\hline
			Metal & \;\;\makecell{$\hbar\Gamma_\text{m} (\SI{}{\electronvolt})$}\;\; & \;\;\makecell{$\hbar\omega_\text{p} (\SI{}{\electronvolt})$}\;\; & \makecell{$v_\text{\tiny{F}} (10^6\SI{}{\meter\per\second})$} & \makecell{$D_\text{m}$\\$(10^{-4}\SI{}{\meter\squared\per\second})$} \\ 
			\hline
			Au & 0.071  & 9.02  & 1.39  & 8.62  \\ 
			Ag & 0.025  & 8.99  & 1.39  & 9.62 \\ 
			\hline
		\end{tabular}
		\caption{Metal parameters from \cite{raza2015nonlocal}.}
		\label{Table:Params_from_Raza}
	\end{center}
\end{table}

\subsection*{Common parameters used}\vspace{-0.3em}
Throughout this work, we use the set of NV parameters obtained by Albrecht \emph{et al.} in \cite{albrecht2013coupling} for a single NV centre in a nanodiamond at room temperature. This has been done by fitting the NV emission intensity spectrum in air with 8 Lorentzian lines ($n=7$) with scaled amplitudes. These amplitudes $A_k$, phonon energies $\hbar\omega_k$, free-space decay rates $\gamma_k^\text{f}$, and phonon decay rates $\gamma_{k,k-1}$ are presented in Table\ \ref{Table:Params_from_Roland}. We modify the free-space decay rates using equations (\ref{Eq:Decay_near_large_MNPs}) or (\ref{Eq:Decay_near_small_MNPs}) to obtain $\gamma_k$ of the NV centre in the presence of an MNP, as discussed earlier. The energy of the NV zero-phonon line $\hbar\omega_\text{z} = \SI{1.941}{\electronvolt}$ \cite{albrecht2013coupling}, and the dephasing rate between the ground and excited states $\gamma_*=\SI{15}{\tera\hertz}$ \cite{albrecht2013coupling} is used for all transitions. 

In \cite{albrecht2013coupling}, each  $\gamma_k^\text{\tiny{f}}$ is obtained by scaling the total decay rate $\gamma_\text{tot}\sim 1/\SI{29}{\nano\second}$ (for NV in nanodiamond) such that  $\gamma_k^\text{\tiny{f}} = \varepsilon_k\cdot\gamma_\text{tot}$ where $\varepsilon_k = A_k\big/\sum_k A_k$. It is noteworthy that the effective excited state lifetime in air (or free-space) ${1/\sum \gamma_k^\text{\tiny{f}}}$ resulting from the decay rates reported by Albrecht \emph{et al}.\ in \cite{albrecht2013coupling} is quite close to the optical excited state ($^3E$) lifetime measurement for a single NV centre in nanodiamond $\sim\SI{25}{\nano\second}$ reported by Beveratos \emph{et al}.\ in \cite{beveratos2001nonclassical}. It is observable that optical excited state ($^3E$) lifetimes in nanodiamond crystals that are considerably smaller than the fluorescence wavelength approximately double in comparison to NV centres in bulk diamond.  This change is has been attributed to the reduction of the radiative emission rate induced by the decrease of the effective refractive index of the medium surrounding the NV centre \cite{doherty2013nitrogen}.

The absolute angle-averaged optical dipole moment element for the NV centre is obtained from \cite{alkauskas2014first} as $\mu_{e\leftrightarrow g}\sim \SI{5.2}{D}$. In our model, we assume $\SI{5.2}{D} \sim \mu_0$ (the dipole moment element that corresponds to the $|e_j\rangle\leftrightarrow|g_0\rangle$ transition for both $j=0$ and $1$). We then estimate the scaled dipole moment elements for other optical transitions $|e_j\rangle\leftrightarrow|g_k\rangle$ as $\mu_k=\sqrt{\varepsilon_k/\varepsilon_0}\mu_0$, such that $\gamma_k^\text{\tiny{f}} \propto |\mu_k|^2$ for each transition (as required by both Fermi's golden rule and Einstein A coefficient for a generic emitter \cite{fox2006quantum, carmichael1999statistical, premaratne2021theoretical}). 

We estimate the total nonradiative decay rate between excited levels as $\gamma_\text{e}\sim\frac{1}{n}\left(\sum_k \gamma_{k,k-1}\right)* n_\text{e}$, where $n_\text{e}$ is the expected number of phonons between $|e_1\rangle$ and $|e_0\rangle$ obtained assuming that the average energy of a phonon in the ground and excited states are similar.

The positive frequency amplitude of the externally incident field is $E_0=\SI{30e4}{\volt\per\meter}$ (such that the resulting Rabi frequencies are in the $\SI{}{\giga\hertz}$ range). The refractive indices of air, water, and PMMA were taken as $n_\text{b}\approx 1$, $n_\text{b}\approx 1.33$, and $n_\text{b}\approx 1.495$, respectively. The refractive index of diamond, $n_\text{\tiny{D}}\approx 2.4$.

{Dielectric permittivity of the MNP $\epsilon_\text{m}$ was obtained by interpolating the tabulations by Johnson and Christy \cite{fox2006quantum} for both Au and Ag. The bound electron response for a given angular frequency $\omega$ was obtained using the relationship, $\epsilon_\text{core}(\omega) = \epsilon_\text{m}(\omega) + \omega_\text{p}^2\big/(\omega^2 + i\omega\Gamma_\text{m})$ \cite{raza2015nonlocal}, where $\omega_\text{p}$ is the bulk plasma frequency of the metal. Additionally, the parameters in Table \ref{Table:Params_from_Raza} obtained from \cite{raza2015nonlocal} were used when modelling the Au and AgNPs.}

The values of any other parameters used (MNP radius $r_\text{m},$ NV-MNP centre separation $R$, the submerging medium considered, and the orientation parameter $s_\alpha$) will be presented alongside each set of results, separately.

\begin{figure*}[t!]
	\includegraphics[width=0.78\textwidth]{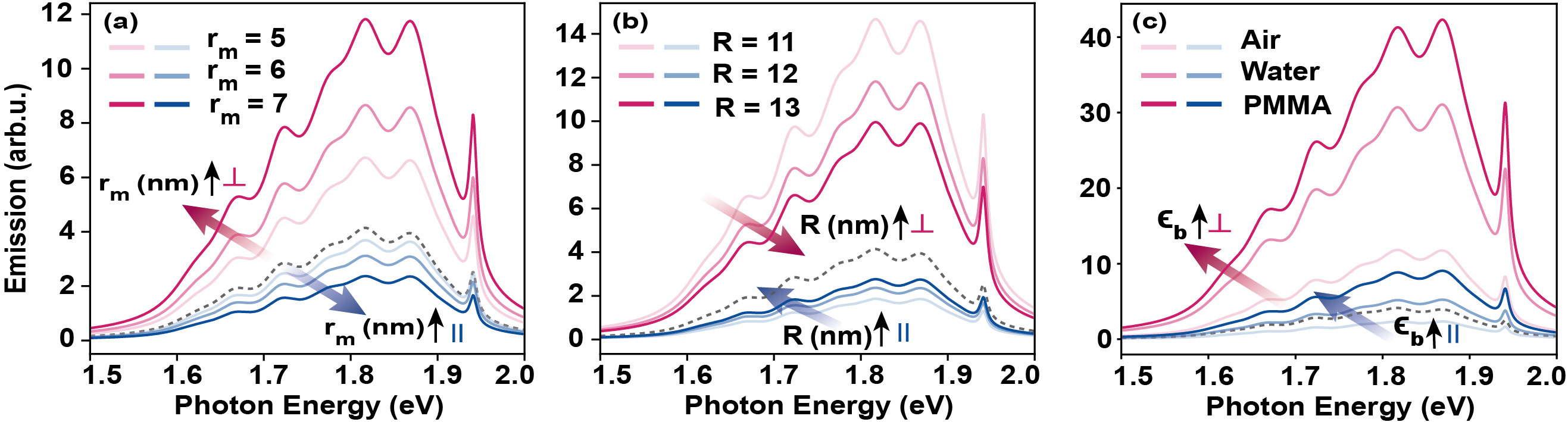}
	\centering
	\caption{(a) MNP radius ($r_\text{m}$) dependence of total near-field NV emission in the presence of small AuNPs of radii 5, 6 and $\SI{7}{\nano\meter}$ at $\SI{12}{\nano\meter}$ centre separation, in air. (b) MNP centre separation ($R$) dependence of NV emission in the presence of a $\SI{7}{\nano\meter}$ radius AuNP in air. (c) Submerging medium dependence of NV emission in the presence of a $\SI{7}{\nano\meter}$ radius AuNP at a $\SI{12}{\nano\meter}$ centre separation. The red and blue shaded curves in all three subplots correspond to NVs in NV$^\perp$MNP ($\perp$) and NV$^\parallel$MNP ($\parallel$) configurations, respectively. The dashed reference line corresponds to the emission intensity of the isolated NV centre in air. All curves are normalized by the area of the respective reference curve. Illumination is at the free-space wavelength $\SI{532}{\nano\meter}$. \label{Fig:Small_MNPs_suppMat}}
\end{figure*}
 
\subsection*{Generation of sample results}
We generated steady state photon emission intensity spectra for the nonlinearly treated NV$^\perp$MNP and NV$^\parallel$MNP configurations using the following procedure: The NV centre was initiated in its zero-phonon ground state and evolved using the previously outlined piecewise superoperator method to obtain the steady state NV density matrix $\hat{\rho}_\text{\tiny{RF}}$. We then obtained the density matrix-dependent steady state Hamiltonian using (\ref{Eq:RF_Hamiltonian}) and (\ref{Eq:E_tot_plus_rabi_form}). The steady state density matrix-based Hamiltonian and all collapse operators accompanied by their respective rates (arranged in the form $\sqrt{\Gamma_x}\hat{L}_x$) were then input to the \emph{spectrum}() function in QuTiP to obtain the emission correlation spectra for each emission operator $\hat{\sigma}_k$. The total NV emission spectrum in the rotating reference frame was then obtained as the summation of such spectra for all NV emission bands, as outlined in (\ref{Eq:Total_emission_intensity}). The obtained spectra were normalized by the area of the respective isolated NV emission intensity spectrum and shifted into the laboratory reference frame by adding $\hbar\omega_\text{d}$ to the emitted photon energies (in the independent axis). 

The same procedure was followed in the absence of MNP induced electric field components, when generating the isolated NV emission spectra in Fig.\ 1(c) of the main text and the reference curves in all figures.

Sample results generated using the above procedure, for MNP radius ($r_\text{m}$), centre separation ($R$) and submerging medium permittivity ($\epsilon_\text{b}$) dependence of the NV emission intensity in the presence of small MNPs are depicted in Fig.\ \ref{Fig:Small_MNPs_suppMat}. It is evident that the behavioural trends of NV emission in the presence of small MNPs are qualitatively equivalent to those observed in the presence of large MNPs, presented in the main text. 

\providecommand{\noopsort}[1]{}\providecommand{\singleletter}[1]{#1}%
%

\end{bibunit}
\end{document}